\documentclass[twocolumn]{aastex7}

\DeclareUnicodeCharacter{02BC}{\-}
\shortauthors{Yan et al.}
\usepackage{textcomp, gensymb}
\usepackage{booktabs}
\usepackage{soul}
\usepackage{float}
\usepackage{comment}
\usepackage{array}
\usepackage{makecell}
\usepackage{amsmath}
\usepackage{multirow} 
\usepackage{enumitem}

\newcommand{\degdot}{\rlap{.}^\circ}

\defcitealias{Yan_2023}{Y23}
\defcitealias{Nakahara_2018ApJ...854..148N}{N18}

\begin{document}
\title{
Sub-Parsec Acceleration and Collimation of NGC 4261's Twin Jets
}

\author[0009-0003-6680-1628]{Xi Yan}
\affiliation{Xinjiang Astronomical Observatory, CAS, 150 Science 1-Street, Urumqi, Xinjiang, 830011, China}
\email{yanxi@xao.ac.cn}

\author[0000-0003-0721-5509]{Lang Cui}
\affiliation{Xinjiang Astronomical Observatory, CAS, 150 Science 1-Street, Urumqi, Xinjiang, 830011, China}
\affiliation{State Key Laboratory of Radio Astronomy and Technology, A20 Datun Road, Chaoyang District, Beijing, 100101, P. R. China}
\affiliation{Xinjiang Key Laboratory of Radio Astrophysics, 150 Science 1-Street, Urumqi 830011, China}
\email[show]{cuilang@xao.ac.cn}

\author[0000-0001-6906-772X]{Kazuhiro Hada}
\affiliation{Graduate School of Science, Nagoya City University, Yamanohata 1, Mizuho-cho, Mizuho-ku, Nagoya, Aichi 467-8501, Japan}
\affiliation{Mizusawa VLBI Observatory, National Astronomical Observatory of Japan, 2-12 Hoshigaoka, Mizusawa, Oshu, Iwate 023-0861, Japan}
\email{hada@nsc.nagoya-cu.ac.jp}

\author[0000-0003-3079-1889]{S\'andor Frey}
\affiliation{Konkoly Observatory, HUN-REN Research Centre for Astronomy and Earth Sciences, Konkoly Thege Mikl\'os \'ut 15-17, H-1121 Budapest, Hungary}
\affiliation{CSFK, MTA Centre of Excellence, Konkoly Thege Mikl\'os \'ut 15-17, H-1121 Budapest, Hungary}
\affiliation{Institute of Physics and Astronomy, ELTE Eötvös Loránd University, P\'azm\'any P\'eter s\'et\'any 1/A, H-1117 Budapest, Hungary}
\email{frey.sandor@csfk.org}

\author[0000-0002-7692-7967]{Ru-sen Lu}
\affiliation{Shanghai Astronomical Observatory, Chinese Academy of Sciences, 80 Nandan Road, Shanghai 200030, China}
\affiliation{State Key Laboratory of Radio Astronomy and Technology, A20 Datun Road, Chaoyang District, Beijing, 100101, P. R. China}
\affiliation{Max-Planck-Institut für Radioastronomie, Auf dem Hügel 69, D-53121 Bonn, Germany}
\email{rslu@shao.ac.cn}

\author[0000-0002-1908-0536]{Liang Chen}
\affiliation{Shanghai Astronomical Observatory, Chinese Academy of Sciences, 80 Nandan Road, Shanghai 200030, China}
\email{chenliang@shao.ac.cn}

\author[0009-0005-8912-4804]{Wancheng Xu}
\affiliation{Xinjiang Astronomical Observatory, CAS, 150 Science 1-Street, Urumqi, Xinjiang, 830011, China}
\affiliation{College of Astronomy and Space Science, University of Chinese Academy of Sciences, No.1 Yanqihu East Road, Beijing 101408, China}
\email{xuwancheng@xao.ac.cn}

\author[0009-0005-4034-1373]{Elika P. Fariyanto}
\affiliation{Department of Astronomy, Graduate School of Science, The University of Tokyo, 7-3-1 Hongo, Bunkyo-ku, Tokyo 113-0033, Japan}
\affiliation{National Astronomical Observatory of Japan, 2-21-1 Osawa, Mitaka, Tokyo 181-8588, Japan}
\email{fariyanto-elika-prameswari@g.ecc.u-tokyo.ac.jp}

\author[0000-0001-6947-5846]{Luis C. Ho}
\affiliation{Kavli Institute for Astronomy and Astrophysics, Peking University, Beijing 100871, China}
\affiliation{Department of Astronomy, School of Physics, Peking University, Beijing 100871, China}
\email{lho.pku@gmail.com}

\correspondingauthor{Lang Cui}

\begin{abstract}
We report the first robust evidence for a co-spatial sub-parsec acceleration and collimation zone (ACZ) in the twin jets of the nearby low-luminosity active galactic nucleus (LLAGN) NGC\,4261. This result is derived from multifrequency Very Long Baseline Array imaging, combined with the frequency-dependent properties of the radio core (core shift and core size) and jet kinematics determined from the jet-to-counterjet brightness ratio. By applying multiple analysis methods and incorporating results from the literature, we identify a parabolic-to-conical structural transition in both the jet and counterjet, with the transition occurring at $(1.23\pm0.24)$\,pc or $(8.1\pm1.6)\times10^3\,R_{\rm s}$ (Schwarzschild radii) for the jet and $(0.97\pm0.29)$\,pc or $(6.4\pm1.9)\times10^3\,R_{\rm s}$ for the counterjet. We also derive the jet velocity field at distances of $\sim (10^3-2\times10^4)\,R_{\rm s}$. While local kinematic variations are present, the jet shows an overall acceleration to relativistic speeds from $\sim 10^3$ to $\sim8\times10^3\,R_{\rm s}$, with a maximum Lorentz factor of $\Gamma_{\rm max} \approx 2.6$. Beyond this region, the jet gradually decelerates to sub-relativistic speeds. These findings support the existence of a sub-parsec-scale ($\lesssim 1.5$\,pc) ACZ in NGC\,4261, where the jet is accelerated via magnetic-to-kinetic energy conversion while being confined by external pressure. A brief comparison with M\,87 suggests that the ACZ in NGC\,4261 may represent a scaled-down analogue of that in M\,87. These results point towards a potential diversity in jet ACZ properties, emphasizing the importance of extending such studies to a broader AGN population to elucidate the physical mechanisms at play.
\end{abstract}


\section{Introduction} \label{sec: Introduction}
The question of how relativistic jets in active galactic nuclei (AGNs) are launched, accelerated, and collimated is a long-standing fundamental issue in astrophysics. From a theoretical point of view, these jets are powered by extracting energy from the inner accretion disk around a spinning supermassive black hole (SMBH) and/or from the SMBH itself via magnetohydrodynamic (MHD) processes(\citealt{blandford_1977MNRAS.179..433B}; \citealt{Komissarov_2009MNRAS.394.1182K}; see \citealt{Blandford_2019ARA&A..57..467B} for a review). As the jet expands, the magnetic energy stored at its base is gradually converted into kinetic energy, which accelerates the jet to relativistic speeds. Meanwhile, the accelerating jet can be effectively collimated by external pressure from its ambient medium \citep[e.g.,][]{Lyubarsky_2009ApJ...698.1570L}. In particular, the processes of acceleration and collimation are causally connected and occur simultaneously, which is known as the acceleration and collimation zone (ACZ).

Very long baseline interferometry (VLBI), thanks to its high spatial resolution, provides the best opportunity to study the physical processes that occur in the ACZs of AGN jets (e.g., \citealt{Boccardi_2017A&ARv..25....4B}). M\,87 represents the most well-studied one (see a recent review by \citealt{Hada_2024}).
The significant progress made in M\,87 has consequently spurred active research on ACZs of other AGNs. 

In terms of jet collimation, this aspect has been widely explored in both low- and high-luminosity radio sources \citep[e.g.,][]{Tseng_2016ApJ...833..288T,Boccardi_CygnusA_2016A&A...585A..33B,Boccardi_3C264_2019A&A...627A..89B,Traianou_2020A&A...634A.112T,Casadio_2021A&A...649A.153C}. For most of the well-studied sources, a parabolic-to-conical transition in the jet collimation profile has been clearly observed. However, there are some exceptions, such as 3C\,84 \citep{Nagai_2014ApJ...785...53N,Giovannini_2018NatAs...2..472G}, Cygnus\,A \citep{Nakahara_CygnusA_2019ApJ...878...61N}, and NGC\,1052 \citep{Nakahara_NGC1052_2020AJ....159...14N,Baczko_2022A&A...658A.119B,Baczko_2024A&A...692A.205B}. Furthermore, it is interesting to note that the location of the break in the parabolic/cylindrical-to-conical transition does not always coincide with the Bondi radius ($r_{\rm Bondi}$) of the central SMBH \citep[e.g.,][]{boccardi_NGC315_2021A&A...647A..67B,Okino_2022ApJ...940...65O,Burd_2022A&A...660A...1B}.

Despite ongoing advancements in our understanding of the jet collimation, observational studies that focus on the detailed acceleration processes of AGN jets remain insufficient. 
Until now, only four AGN jets, to the best of our knowledge, provide solid evidence of concurrent acceleration and collimation. These include two low-luminosity AGNs (LLAGNs), M\,87 \citep{Asada_2012ApJ...745L..28A,Park_kinematics_2019ApJ...887..147P} and NGC\,315 \citep[e.g.,][]{park_2021ApJ...909...76P,boccardi_NGC315_2021A&A...647A..67B,Ricci_2022A&A...664A.166R,Ricci_2025A&A...693A.172R}, a narrow-line Seyfert 1 (NLS1) galaxy, 1H 0323+342 \citep{Hada_1H0323_2018ApJ...860..141H}, and a flat-spectrum radio quasar (FSRQ), 1928+738 \citep{Yi_2024A&A...688A..94Y}.

Motivated by this limitation, we have selected NGC\,4261 (3C\,270), a nearby LLAGN (see \citealt{HO_2008ARA&A..46..475H} for a review), as the focus of our study. This source is particularly noteworthy for several reasons. First, its relatively close distance \citep[$(31.6\pm2.8)$\,Mpc;][]{Tonry_2001ApJ...546..681T,Cappellari_2011MNRAS.413..813C} and the existence of a SMBH at its center \citep[$M_{\rm BH}= (1.62\pm0.04)\times10^{9}M_{\sun}$; e.g.,][]{ruffa_2023MNRAS.522.6170R} allow for a spatial resolution of 1\,mas $\sim$ 0.15\,pc $\sim$ 988 Schwarzschild radii ($R_{\rm s}$). Hence, NGC\,4261 hosts one of the most promising SMBHs for future submillimeter VLBI imaging by Event Horizon Telescope (EHT) or next-generation EHT (ngEHT) \citep[e.g.,][]{Zhang_2025ApJ98541Z}. Moreover, the location of the central SMBH has been precisely determined by measuring the core shift in both the jet and counterjet by \citet{haga2015ApJ80715H}. Lastly, the jet of NGC\,4261 is inclined at a relatively large viewing angle in the range of $54^{\circ} \lesssim \phi_{\rm view} \lesssim 84^{\circ}$, with the most probable value being ${71^{\circ}} \pm 2^{\circ}$ (1$\sigma$) (\citealt{Yan_2024evn..conf...67Y}). Consequently, NGC\,4261 provides a valuable opportunity to study the physical properties of its two-sided jets not only on the kiloparsec scales \citep[e.g.,][]{Birkinshaw_1985ApJ...291...32B,Kolokythas_2015MNRAS.450.1732K} but also on the VLBI or sub-parsec scales \citep[e.g.,][]{Jones_1997ApJ...484..186J}.

The jet geometry of NGC\,4261 has been explored by \citet{Nakahara_2018ApJ...854..148N} (hereafter \citetalias{Nakahara_2018ApJ...854..148N}) and \citet{Yan_2023} (hereafter \citetalias{Yan_2023}). These studies reveal a transition in the jet collimation profile from parabolic to conical shape on the subparsec scales of $\lesssim 1$\,pc (deprojected). Interestingly, this break location is significantly smaller than the Bondi radius of the central SMBH ($r_{\rm Bondi} \approx 99.2$\,pc; \citealt{Balmaverde_2008A&A...486..119B}; \citetalias{Yan_2023}). On the other hand, a preliminary analysis of jet kinematics by \citetalias{Yan_2023}, using four-epoch VLBI data at 15\,GHz, tentatively suggests that the jet acceleration zone may coincide with its collimation zone.

\begin{deluxetable*}{cccccccccccc}[htbp!]
\renewcommand\arraystretch{1.3}
\tabletypesize{\footnotesize}
\tablecaption{Summary of the Observations and Data of NGC\,4261 
\label{tab: NGC4261_observation_summary}}
\tablehead{ \colhead{Code} & \colhead{$\nu$} & \colhead{Epoch} & \colhead{Array} & \colhead{Pol.} & \colhead{B.W.} & \colhead{$T_{\rm int}$} & \colhead{$\Theta_{\rm maj} \times \Theta_{\rm min}$, PA} & \colhead{ $I_{\rm peak}$} & \colhead{$I_{\rm rms}$} & \colhead{$S_{\rm tot}$} & \colhead{Publ.}\\
& (GHz) & & & & (MHz) & (min) & (mas $\times$ mas, deg) & \multicolumn{2}{c}{(mJy\,beam$^{-1}$)} & (mJy)\\
(1) & (2) & (3) & (4) & (5) & (6) & (7) & (8) & (9) & (10) & (11) & (12)
}
\startdata
\multirow{2}{*}{BW011}  & 1.7  & \multirow{2}{*}{1995/04/01}  &  \makecell{VLBA \\ (MK$^a$, SC$^a$)} & LCP & 64 & $\sim235$ & $9.75\times5.14, 10.6$& 78.6 & 0.25 & $191\pm19$ & \multirow{2}{*}{\makecell{(a),(b)\\(c),(d)}}\\
        & 8.4  &  & VLBA & RCP & 64 & $\sim75$ & $1.92\times0.86, 1.7$ & 121.1 & 0.30 & $350\pm35$\\
\hline
\multirow{2}{*}{BJ028}  & 8.4 & 1999/02/26 & VLBA & RCP & 64 & $\sim475$ & $1.69\times0.73, -3.7$ & 99.8 & 0.10  & $299\pm30$ & \multirow{2}{*}{(d)} \\
       & 8.4 & 1999/10/21 & VLBA, $-$BR & RCP & 64 & $\sim 460$ & $2.06\times0.70, -8.8$ & 124.3 & 0.11 & $332\pm33$ & \\
\hline
\multirow{4}{*}{BS094b} & 1.4 & \multirow{4}{*}{2002/08/29} & \multirow{4}{*}{\makecell{VLBA, $-$BR}} & Dual & 32 & $\sim60$  & $10.99\times4.54, -3.7$ & 55.6 & 0.30 & $150\pm15$ & \multirow{4}{*}{...} \\
       & 2.3 &  &  & RCP & 16 & $\sim60$  & $7.01\times2.85, -4.8$  & 97.1 & 0.45 & $210\pm21$ & \\
       & 5.0 &  &  & LCP & 32 & $\sim60$  & $3.13\times1.26, -4.7$  & 113.5 & 0.30 & $300\pm30$ & \\
       & 8.4 &  &  & RCP & 16 & $\sim60$  & $1.98\times0.79, -4.3$  & 145.8 & 0.55 & $317\pm32$ & \\
\hline
\multirow{2}{*}{BA064} & 4.8 & \multirow{2}{*}{2002/12/14} & \multirow{2}{*}{\makecell{VLBA \\ (FD, KP, LA, \\ NL, OV, PT)}}  & Dual & 64 & $\sim25$  & $9.02\times4.27, -20.7$  & 202.0 & 0.49 & $304\pm30$ & \multirow{2}{*}{...} \\
       & 8.3 & &  & Dual & 64 & $\sim25$  & $5.40\times2.59, -20.6$  & 221.8 & 0.46 & $337\pm34$ & \\
\hline
\multirow{3}{*}{BS094c}  & 15 & \multirow{3}{*}{2003/06/28}  &  \multirow{3}{*}{VLBA} & \multirow{3}{*}{LCP} & \multirow{3}{*}{32} & $\sim60$  & $0.97\times0.43, -0.2$  & 125.1 & 0.81 & $307\pm31$ & \multirow{3}{*}{\makecell{(e) \\ (f) \\ (g)}} \\
       & 22 &  &  &  & & $\sim60$  & $0.66\times0.26, -1.7$ & 122.4 & 1.30 & $283\pm28$ & \\
       & 43 &  &  &  & & $\sim60$  & $0.35\times0.16, 2.7$  & 135.7 & 2.30 & $271\pm27$ & \\
\hline
\multirow{4}{*}{BS094d} & 1.4 & \multirow{4}{*}{2003/07/05} &  \multirow{4}{*}{VLBA} & Dual & 32 & $\sim60$  & $9.70\times4.63, -1.2$  & 54.1 & 0.50 & $162\pm16$ & \multirow{4}{*}{\makecell{(e) \\ (f) \\ (g)}} \\
       & 2.3 &  &  & RCP & 16 & $\sim60$  & $6.15\times2.95, -2.4$  & 98.6 & 0.50 & $222\pm22$ & \\
       & 5.0 &  &  & LCP & 32 & $\sim60$  & $2.82\times1.33, -4.1$ & 122.2 & 0.26 & $296\pm30$ & \\
       & 8.4 &  &  & RCP & 16 & $\sim60$  & $1.77\times0.80, -3.0$ & 121.5 & 0.50 & $299\pm30$ & \\
\enddata
\tablecomments{
Column\,(1): project code. Note that BS094b/c/d are phase-referencing observations \citep[see details in][]{haga2015ApJ80715H}.
Column\,(2): frequency. 
Column\,(3): observation date. 
Column\,(4): participating stations (stations not participating are marked with a minus sign). 
Column\,(5): polarization. LCP -- left circular polarization, RCP -- right circular polarization, Dual -- LCP and RCP.
Column\,(6): bandwidth.
Column\,(7): on-source integration time.
Column\,(8): FWHM of the major and minor axes of the elliptical Gaussian synthesized beam, along with the position angle (PA) of the major axis.
Columns\,(9)--(11): peak intensity, rms noise level ($\sigma$), and total flux density of the CLEAN image.
Column\,(12): publications in which the data have been used: 
(a) \citet{Jones_1997ApJ...484..186J}; 
(b) \citet{Jones_2000ApJ...534..165J}; 
(c) \citet{Jones_2001ApJ...553..968J}; 
(d) \citet{piner_2001AJ122.2954P}; 
(e) \citet{haga2015ApJ80715H}; 
(f) \citetalias{Nakahara_2018ApJ...854..148N}; 
(g) \citetalias{Yan_2023}. 
Note that BS094c/d have been used to measure the core shift and jet collimation profile (see \citealt{haga2015ApJ80715H}; \citetalias{Nakahara_2018ApJ...854..148N}; \citetalias{Yan_2023}).
\flushleft $^{a}$Fringe detection is marginal \citep[see also][]{Jones_1997ApJ...484..186J}.
}
\end{deluxetable*}

In the high-energy regime, observations have revealed the X-ray jet and counterjet on scales of a few kiloparsecs, suggesting a possible synchrotron origin for the X-ray emission from the jets in NGC\,4261 \citep{Sambruna_2003ApJ...586L..37S, Gliozzi_2003A&A...408..949G, Zezas_2005ApJ...627..711Z, Worrall_2010MNRAS.408..701W}. In particular, observations from the Large Area Telescope (LAT) on board the \textit{Fermi} satellite (\textit{Fermi}-LAT) have identified NGC\,4261 as a $\gamma$-ray emitter \citep{Menezes_2020MNRAS.492.4120D}. Broadband spectral energy distribution modeling further suggests that $\gamma$-ray radiation could originate from both the subparsec-scale (for $\nu \lesssim 0.6$\,GeV) and kiloparsec-scale (for $\nu \gtrsim 0.6$\,GeV) jets \citep{Tomar_2021ApJ...919..137T}.

In this paper, we report a VLBI study of jet acceleration and collimation in NGC\,4261, using archival Very Long Baseline Array (VLBA) data. Section~\ref{sec: Observations} presents historical observations and data reduction. The results are shown in Section~\ref{sec: Results}, followed by a discussion in Section~\ref{sec: Discussion}. We conclude with a summary of our main findings in Section~\ref{sec: Summary}. Throughout this work, we adopt a weighted mean viewing angle of $\phi_{\rm view} = 68^{\circ}\pm 4^{\circ}$ for the jet, derived from the values of ${63^{\circ}}\pm3^{\circ}$ and ${71^{\circ}}\pm2^{\circ}$ reported by \citet{piner_2001AJ122.2954P} and \citet{Yan_2024evn..conf...67Y}, respectively. We also define the spectral index ($\alpha$) as $S\propto\nu^{+\alpha}$. For a distance of $31.6$\,Mpc to NGC\,4261, 1\,$c$ (the speed of light) corresponds to approximately $2$\,mas\,yr$^{-1}$ apparent proper motion.

\begin{figure*}
\begin{center}
    \includegraphics[width=0.32\linewidth]{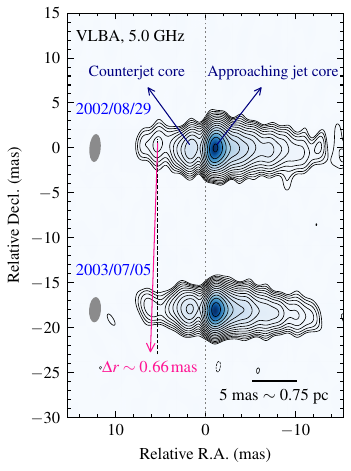}
    \hspace{0.25cm} 
    \includegraphics[width=0.31\linewidth]{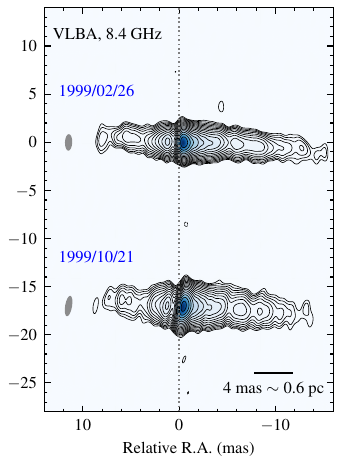} 
    \hspace{0.25cm}
    \includegraphics[width=0.315\linewidth]{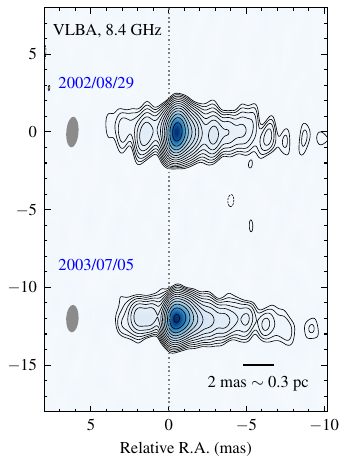}
    \includegraphics[width=0.34\linewidth]{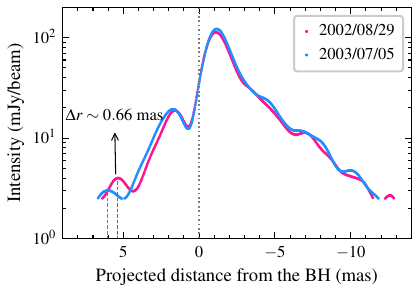}
    \includegraphics[width=0.32\linewidth]{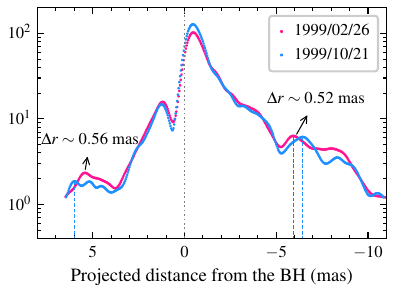}
    \includegraphics[width=0.32\linewidth]{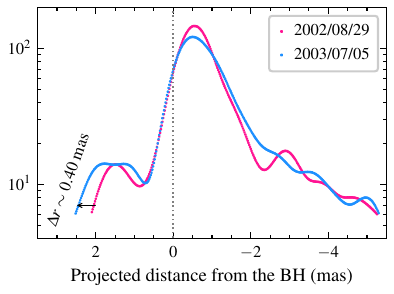} 
\caption{\textit{Top panels:} uniformly weighted CLEAN images of the NGC\,4261 jet observed with VLBA at 5.0 and 8.4\,GHz. Contours start from the 3$\sigma$ image rms noise (see Table~\ref{tab: NGC4261_observation_summary}) and increase by a factor of $\sqrt{2}$. The synthesized beam is shown in the left of each contour. We shift the putative location of the central SMBH to the image center based on the core shifts measured by \citet{haga2015ApJ80715H}, as marked by the gray dotted line in each plot. \textit{Bottom panels:} radial intensity profile corresponding to each plot in the top panel. All measurements are taken with S/Ns $\geq$ 10. These profiles are used to analyze the possible motions of the twin jets in NGC\,4261 (see Section~\ref{sec: Results_Jet_velocity_field} and Appendix~\ref{Appendix: Jet_motions_derived_from_radial_intensity_profiles}). Note that the 5.0 and 8.4\,GHz images clearly reveal the cores of both the approaching jet and counterjet (as indicated in the 5.0\,GHz image), consistent with the identification by \citet{haga2015ApJ80715H}. These detections allow them to measure the core shifts in both jets. 
\label{fig: NGC4261_jet_and_cjet_motions}}
\end{center}
\end{figure*}

\section{Observations and Data Reduction} \label{sec: Observations}
In this study, we used archival VLBA data observed at frequencies between 1.4 and 43\,GHz. The visibility data were obtained from the U.S. National Radio Astronomy Observatory (NRAO) Data Archive\footnote{\url{https://data.nrao.edu/}}. The details of these observations are summarized in Table~\ref{tab: NGC4261_observation_summary}. As noted in column (12), some of the data have not yet been published in the literature.

We calibrated the data following standard procedures with the NRAO Astronomical Image Processing System \citep[{\tt AIPS};][]{Greisen2003}. We first corrected for the digital sampler bias using the auto-correlation spectra. Then, parallactic angle variations, residual ionospheric delays, and Earth orientation parameters were corrected or updated. After removing the instrumental delay and phase offset for each intermediate frequency, we performed global fringe fitting on the target and calibrators to derive the residual phase, delay, and rate solutions. In cases where residual solutions for NGC\,4261 cannot be determined due to low signal-to-noise ratios (S/Ns) (i.e., the BS094b/c/d data sets, see Table~\ref{tab: NGC4261_observation_summary}), global fringe fitting was only run on the phase-referencing calibrator J1222+0413 (separated by $1\degdot$77 on the sky). The residual phase, delay, and rate solutions derived from the calibrator were then interpolated into consecutive scans of NGC\,4261 (see details in \citealt{haga2015ApJ80715H}). We performed a priori amplitude calibration with opacity corrections using system temperatures and antenna gain curves. The bandpass corrections were applied using the bright calibrators. Finally, we performed iterative CLEAN and phase/amplitude self-calibration, followed by imaging using the Caltech {\tt DIFMAP} package \citep{Shepherd_Difmap_1997ASPC..125...77S}.

\section{Data Analysis and Results} \label{sec: Results}
Figure~\ref{fig: NGC4261_jet_and_cjet_motions} shows the representative CLEAN images of NGC\,4261, while the remaining images are provided in Figures~\ref{fig: NGC4261_1.4-8.4GHz_images} and \ref{fig: NGC4261_15-43GHz_images} in Appendix~\ref{Appendix: CLEAN_Images} (see Table~\ref{tab: NGC4261_observation_summary} for image parameters). All these images (except at 22 and 43\,GHz) clearly reveal the two-sided jets along the east--west direction, with the western side being the approaching jet and the eastern side the counterjet. At 5, 8.4, and 15\,GHz, an emission gap is evidently visible in the innermost counterjet, which is attributed to free-free absorption (FFA) by the foreground ionized gas from the accretion disk (e.g., \citealt{Jones_1997ApJ...484..186J,haga2015ApJ80715H}; see also Appendix~\ref{Appendix: Spectral_index_maps}). At 22 and 43\,GHz, the counterjet cannot be detected due to the limited sensitivity of these data. However, recent high-sensitivity observations have successfully detected twin jets at both of these high frequencies (\citealt{Sawada-Satoh_2023PASJ...75..722S}; \citetalias{Yan_2023}). 

In Figure~\ref{fig: NGC4261_jet_and_cjet_motions}, we also show the longitudinal radiation intensity of the twin jets as a function of distance from the central black hole (i.e., the radial intensity profile), from which the proper motions of both the jet and counterjet may be inferred (see Section~\ref{sec: Results_Jet_velocity_field}). Thanks to the clear detection of the counterjet, we are able to study the jet collimation on both sides and probe the jet acceleration by using the jet-to-counterjet brightness ratio and spectral index, as detailed below.

\subsection{Jet Collimation Profile} \label{sec: Results_jet_collimation}
While the collimation profile of the NGC\,4261 jet has been studied by \citetalias{Nakahara_2018ApJ...854..148N} and \citetalias{Yan_2023}, it should be noted that the jet geometry transition, suggested to occur on subparsec scales of $\lesssim 1$\,pc, is derived from the same VLBI data sets observed in 2003 (i.e., BS094c/d in Table~\ref{tab: NGC4261_observation_summary}). In this study, we expand these studies by incorporating additional data sets, including the 1.7/8.4\,GHz data from project BW011 in 1995 and the two-epoch 8.4\,GHz data from project BJ028 in 1999. These data sets offer the advantage of broader bandwidth and/or longer integration time on the source than the BS094c/d data sets (see Table~\ref{tab: NGC4261_observation_summary}). Furthermore, we also include the 5\,GHz data from project BS094b in 2002.  

To robustly determine the jet collimation profile of NGC\,4261, we adopted multiple data analysis methods: (1) deriving the deconvolved jet width based on the Gaussian fit to the transverse intensity profile of the jet, as commonly used before \citep[e.g.,][]{Asada_2012ApJ...745L..28A,Hada_2016ApJ...817..131H,Tseng_2016ApJ...833..288T,Nakahara_2018ApJ...854..148N,Nakahara_CygnusA_2019ApJ...878...61N,Nakahara_NGC1052_2020AJ....159...14N,park_2021ApJ...909...76P,Shang_2025ApJ...986..198S}; (2) estimating the jet width from the model-fitted size of circular Gaussian components \citep[e.g.,][]{boccardi_NGC315_2021A&A...647A..67B,Kravchenko_2025}; (3) exploring the innermost jet geometry using the multifrequency properties of the VLBI core \citep[i.e., the core shift and core size; e.g.,][]{Hada_2013_M87,Nakamura_2013ApJ...775..118N,park_2021ApJ...909...76P}. In the following, we provide detailed descriptions of these methods and present the corresponding results.

\subsubsection{Deconvolved Jet Width} \label{sec: Results_Deconvolved_jet_width}
We determined the deconvolved jet width of NGC\,4261 following the approach described in \citet{park_2021ApJ...909...76P}. First, we repositioned the putative location of the SMBH to the center of each CLEAN image, based on the well-constrained core shifts of \citet{haga2015ApJ80715H}, as indicated in the CLEAN images shown in Figure~\ref{fig: NGC4261_1.4-8.4GHz_images}. We restored each image using a circular beam with a size equal to the major axis of the synthesized beam ($\Theta_{\rm maj}$ in Table~\ref{tab: NGC4261_observation_summary}). The transverse intensity profile of the jet at each distance from the black hole was then extracted through a pixel-based analysis. By fitting a Gaussian function to the transverse intensity profile, we obtained the full width at half maximum (FWHM; $\Theta_{\rm fit}$), from which the deconvolved jet width was calculated as $W = \sqrt{\Theta_{\rm fit}^{2} - \Theta_{\rm maj}^{2}}$. Measurements were retained only if (1) the amplitude of the fitted Gaussian function exceeded 15$\sigma$ image rms noise level, and (2) $\Theta_{\rm fit} > \Theta_{\rm maj}$. Considering the complex blending effect in the core region, we discarded measurements taken at separations from the observed core smaller than the restoring beam size ($\Theta_{\rm maj}$). Finally, the deconvolved widths were binned by distance from the central engine, using a bin size of $\Theta_{\rm maj}/2$. For each bin, the mean width was adopted as the representative jet width, assuming an uncertainty of $\Theta_{\rm maj}/10$.

\begin{figure*}[htbp!]
\vspace{-1cm}
\begin{center}
    \includegraphics[width=0.9\linewidth]{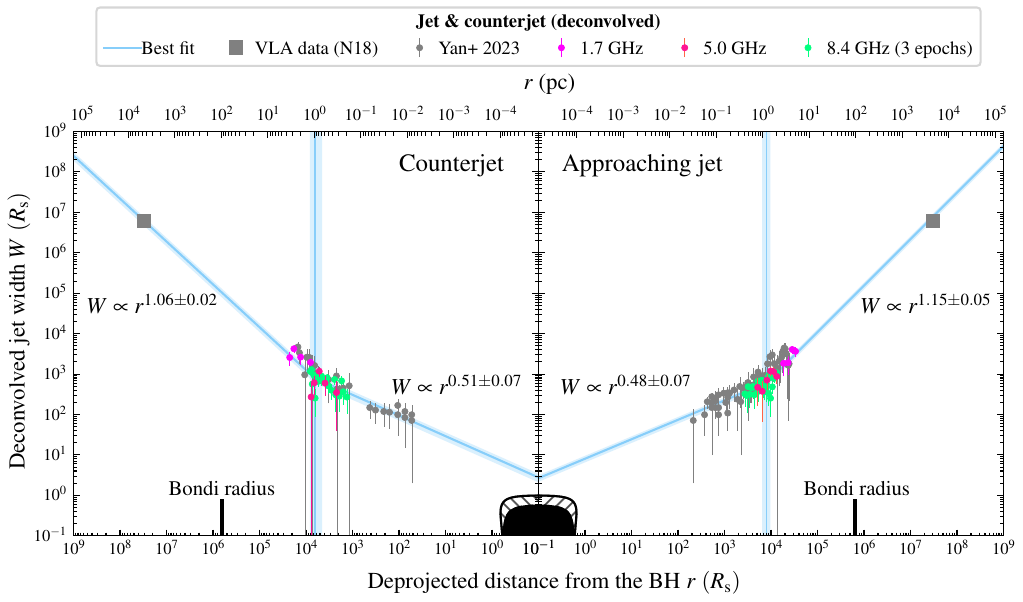}
\caption{Deconvolved widths of the jet (right) and counterjet (left) in NGC\,4261 as functions of deprojected distance from the central black hole, adopting a jet viewing angle of $\phi_{\rm view} = 68^{\circ}\pm 4^{\circ}$ (see Section~\ref{sec: Introduction}). Gray data points are adopted from \citetalias{Yan_2023}. To show the evolution of the jet width on the kiloparsec scales, a VLA measurement from \citetalias{Nakahara_2018ApJ...854..148N} is also included. The best-fit broken power-law functions are overlaid, with their parameters listed in Table~\ref{tab: NGC4261_broken_power-law_fit}. The shaded light blue regions represent the $1\sigma$ uncertainties in the fitted parameters. The break locations are $(1.23 \pm 0.24)$\,pc $\approx (8.1\pm1.6)\times10^3\,R_{\rm s}$ for the approaching jet and $(0.97 \pm 0.29)$\,pc $\approx (6.4\pm1.9)\times10^3\,R_{\rm s}$ for the counterjet, both of which are significantly smaller than the Bondi radius of the central engine, $r_{\rm Bondi} \approx 99.2$\,pc $\approx 6.5\times10^{5}\,R_{\rm s}$. In both panels, the filled black area denotes the event horizon of a Kerr black hole with spin parameter $a = 0.99$, while the hatched region represents its ergosphere.}
\label{fig: NGC4261_collimation_SLICE}
\end{center}
\end{figure*}

In Figure~\ref{fig: NGC4261_collimation_SLICE}, we separately present the deconvolved width of the jet and counterjet as a function of deprojected distance from the central black hole ($r$). Our results overlay those reported by \citetalias{Yan_2023}, which were obtained using a similar method. To illustrate the evolution of the jet width on the kiloparsec scales, we include a Very Large Array (VLA) measurement from \citetalias{Nakahara_2018ApJ...854..148N}, corresponding to $W \approx 6\times10^{6}\,R_{\rm s}$ at a deprojected distance of $r \approx 3\times10^{7}\,R_{\rm s}$\footnote{\citetalias{Nakahara_2018ApJ...854..148N} adopted a black hole mass of $4.9\times10^{8}\,M_{\odot}$ \citep{Ferrarese_1996ApJ...470..444F}, which is about 3.3 times smaller than the value used in this study ($1.62\times10^{9}\,M_{\odot}$; see Section~\ref{sec: Introduction}). However, the larger black hole mass adopted here has been independently determined by three separate studies \citep{Boizelle_2021ApJ...908...19B, Sawada-Satoh_2022A&A...664L..11S, ruffa_2023MNRAS.522.6170R}. In applying their results to this work, we have considered this difference.}.

\begin{deluxetable*}{lccccc}[htbp!]
\tablecaption{Results of Broken Power-law Fit to the Jet Width Profile \label{tab: NGC4261_broken_power-law_fit}}
\tablehead{\colhead{} & \colhead{$W_{0}$} & \colhead{$r_{\rm b}$} & \colhead{$k_{\rm u}$}  & \colhead{$k_{\rm d}$} & \colhead{Reduced $\chi^2$} \\
 & ($R_{\rm s}$) & ($\times 10^3 R_{\rm s}$/pc) &  &  &  \\
 & (1) & (2) & (3) & (4) & (5)}
\startdata
Approaching jet & $631\pm117$ & $8.1\pm1.6$/$1.23\pm0.24$ & $0.48\pm0.07$ &  $1.15\pm0.05$ & 0.89\\
Counterjet & $812\pm195$ & $6.4\pm1.9$/$0.97\pm0.29$  & $0.51\pm0.07$ &  $1.06\pm0.02$ & 1.03  \\
\enddata
\tablecomments{
Column\,(1): jet width at the structural transition.
Column\,(2): break location of the width profile in units of $R_{\rm s}$ and parsec (deprojected).
Columns\,(3)--(4): power-law index of the jet width profile upstream ($k_{\rm u}$) and downstream ($k_{\rm d}$) of the break.
Column\,(5): reduced $\chi^2$ value of the fit.
}
\end{deluxetable*}

Because a structural transition in the NGC\,4261 jet has been previously suggested \citepalias{Nakahara_2018ApJ...854..148N, Yan_2023}, we fitted the jet and counterjet width profiles using a broken power-law function as adopted in \citetalias{Nakahara_2018ApJ...854..148N}:
\begin{equation} \label{eq: broken_power_law}
    W(r) = W_0 \, 2^{(k_{\rm u} - k_{\rm d})/s} \left( \frac{r}{r_{\rm b}} \right)^{k_{\rm u}} \left[ 1 + \left( \frac{r}{r_{\rm b}} \right)^s \right]^{(k_{\rm d} - k_{\rm u})/s},
\end{equation}
where $W_0$ is the jet width at the transition point, $r_{\rm b}$ is the break location in the collimation profile, $k_{\rm u}$ and $k_{\rm d}$ are the power-law indices upstream and downstream of the break, respectively, and $s$ is the sharpness parameter of the transition. Following \citetalias{Nakahara_2018ApJ...854..148N}, we fix $s = 10$. The results of the broken power-law fits for the jet and counterjet, with reduced $\chi^2$ (i.e., $\chi^2$/d.o.f.) values of 0.89 and 1.03, respectively, are listed in Table~\ref{tab: NGC4261_broken_power-law_fit} and shown in Figure~\ref{fig: NGC4261_collimation_SLICE}. As seen, the fitted power-law indices ($k_{\rm u}$ and $k_{\rm d}$) for the jet and counterjet are in good agreement and are roughly consistent with those reported in \citetalias{Nakahara_2018ApJ...854..148N} within the uncertainties (see their Table~2).

Moreover, we note that \citetalias{Nakahara_2018ApJ...854..148N} reported break locations of $(0.75 \pm 0.31)$\,pc for the jet and $(0.37 \pm 0.23)$\,pc for the counterjet, although their measurements in the inner regions are relatively sparse. In comparison, our derived break locations are somewhat larger, i.e., $(1.23 \pm 0.24)$\,pc and $(0.97 \pm 0.29)$\,pc for the approaching and receding sides, respectively (see Table~\ref{tab: NGC4261_broken_power-law_fit})\footnote{It is worth noting that the break locations of the jet and counterjet are not identical. Assuming that the intrinsic jet geometry is symmetric on both sides, the observed differences may result from the effects of a circumnuclear obscuring torus and differences in relativistic beaming between the jet and counterjet (despite the relatively large viewing angle of the NGC\,4261 jet).}. This updated constraint benefits from a larger number of measurements and improved resolution in the inner regions \citepalias{Yan_2023}. We note that the two sets of results remain broadly consistent within the uncertainties.

As shown in Figure~\ref{fig: NGC4261_collimation_SLICE}, our results clearly reveal a consistent transition in the width profiles from parabolic to conical geometry for both the approaching and counterjet sides in NGC\,4261, with the transition occurring at $(0.97 \pm 0.29)\,\text{pc}\lesssim r \lesssim (1.23 \pm 0.24)\,\text{pc}$ or $(6.4\pm1.9)\times10^3\,R_{\rm s} \lesssim r \lesssim (8.1\pm1.6)\times10^3\,R_{\rm s}$. These results indicate that jet collimation on both sides is already complete on the sub-parsec scales---considerably smaller than the Bondi radius of the central SMBH ($r_{\rm Bondi} \approx 99.2$\,pc $\approx 6.5\times10^{5}\,R_{\rm s}$; \citealt{Balmaverde_2008A&A...486..119B}; \citetalias{Yan_2023})---as previously noted by \citetalias{Nakahara_2018ApJ...854..148N} and \citetalias{Yan_2023}.

\subsubsection{Model-fitted Jet Width} \label{sec: Results_Model-fitted_jet_width}
Alternatively, the size of the jet components can serve as a good trace of the jet geometry \citep[e.g.,][]{boccardi_NGC315_2021A&A...647A..67B,Kravchenko_2025}. Therefore, we fitted several circular Gaussian components to the data in {\tt DIFMAP} using the {\tt MODELFIT} subroutine and adopted the fitted FWHM as the jet width. The data and component parameters are summarized in Table~\ref{tab: NGC4261_modelfitted_jet_components} of Appendix~\ref{Appendix: Tables}. Following \citet{boccardi_NGC315_2021A&A...647A..67B}, we assumed a 10\% uncertainty in the flux density for each component and set the size uncertainty to be 25\% of the fitted FWHM. Using the obtained radial distance ($r_{\rm fit}$) of the core and each component, along with the core shifts measured by \citet{haga2015ApJ80715H}, we calculated the projected radial distance of each component relative to the central black hole ($r_{\rm to\,BH}$ in Table~\ref{tab: NGC4261_modelfitted_jet_components}). The uncertainty in $r_{\rm to\,BH}$ was estimated as the quadrature sum of $r_{\rm fit}/5$ and the uncertainty of the core shifts \citep[see][]{haga2015ApJ80715H}. Due to resolution limitations, we excluded components with FWHM smaller than half of the minor-axis beam size ($\Theta_{\rm min}/2$ in Table~\ref{tab: NGC4261_modelfitted_jet_components}).

\begin{figure}[htbp!]
\begin{center}
    \includegraphics[width=1\linewidth]{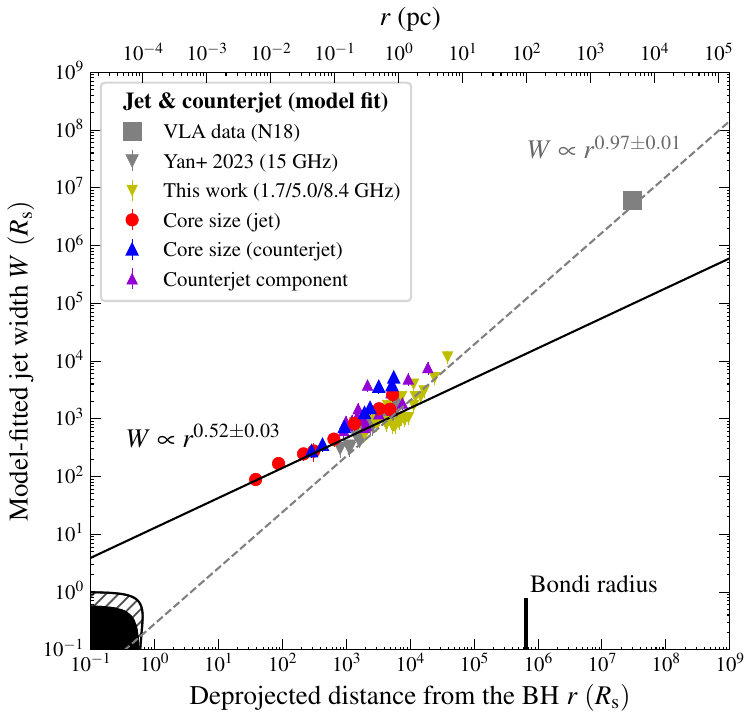}
\caption{Model-fitted widths of the core, jet, and counterjet components in NGC\,4261 as a function of deprojected distance from the central black hole, obtained at 1.7/5.0/8.4\,GHz in this work (see Table~\ref{tab: NGC4261_modelfitted_jet_components} in Appendix~\ref{Appendix: Tables}) and at 15\,GHz by \citetalias{Yan_2023} (4 epochs; see their Table~2). Red points and blue triangles indicate the multifrequency core size of the jet and counterjet, respectively (see Table~\ref{tab: NGC4261_modelfitted_core_size} in Appendix~\ref{Appendix: Tables}). Note that in Figure~\ref{fig: NGC4261_collimation_SLICE}, the jet width is measured using transverse intensity profiles of the jet (see Section~\ref{sec: Results_Deconvolved_jet_width}), whereas in this figure, the jet width is derived from model-fitted FWHMs of the core and jet components (see Section~\ref{sec: Results_Model-fitted_jet_width}).}
\label{fig: NGC4261_collimation_MF}
\end{center}
\end{figure}

\begin{figure*}[ht!]
\begin{center}
    \includegraphics[width=0.24\linewidth]{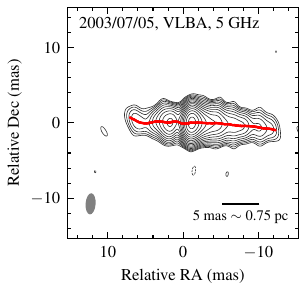}
    \includegraphics[width=0.34\linewidth]{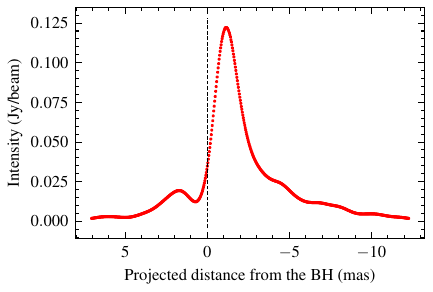}
    \includegraphics[width=0.40\linewidth]{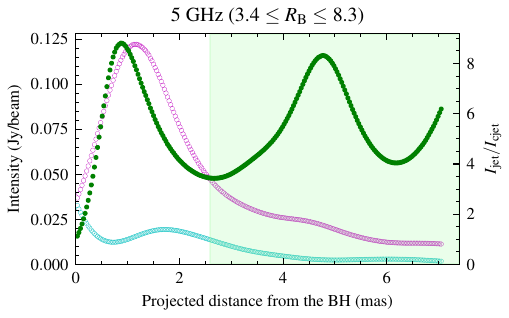}
\caption{A representative 5\,GHz image of NGC\,4261 (left) illustrating how the radial intensity profile (middle) and the jet-to-counterjet brightness ratio (right) are derived. The ridgeline in the CLEAN image represents the maximum intensity along each slice perpendicular to the jet axis. In the right plot, the jet-to-counterjet brightness ratio is constrained by measurements within the green shaded area, considering the complex effects of blending, SSA, and FFA at the jet base (see the text in Section~\ref{sec: Results_measurements_of_BR_and_alpha}).
\label{fig: NGC4261_jet_ridge_line}}
\end{center}
\end{figure*}

As demonstrated in studies of M\,87 and NGC\,315 \citep{Hada_2013_M87, park_2021ApJ...909...76P}, the multifrequency properties of the VLBI core (i.e., core size and core shift) can be used to trace the geometry of the innermost jet. Therefore, we also determined the core size of NGC\,4261 using the data listed in Table~\ref{tab: NGC4261_observation_summary}. Specifically, we modeled the visibility data in {\tt DIFMAP} by fitting a set of circular Gaussian components and identified the brightest component as the core. To better constrain core size, we incorporated additional data from \citetalias{Yan_2023}, which were observed at frequencies of 15\,GHz (4 epochs), 43\,GHz, 44\,GHz, and 88\,GHz. An uncertainty equal to 25\% of the fitted FWHM was adopted for the core size. The results, including the estimated core sizes and core positions relative to the central SMBH (adopted from \citealt{haga2015ApJ80715H}), are summarized in Table~\ref{tab: NGC4261_modelfitted_core_size} of Appendix~\ref{Appendix: Tables}.

We also note that the core shift in the counterjet of NGC\,4261 has also been well determined by \citet{haga2015ApJ80715H}. Following their methodology, we identified the eastern side component closest to the core of the approaching jet as the counterjet core (see Figure~\ref{fig: NGC4261_jet_and_cjet_motions}). The corresponding results are also listed in Table~\ref{tab: NGC4261_modelfitted_core_size} of Appendix~\ref{Appendix: Tables}.

In Figure~\ref{fig: NGC4261_collimation_MF}, we present the model-fitted width for the core, jet, and counterjet as a function of deprojected distance from the central black hole. We also include the model-fitted results from \citetalias{Yan_2023}, based on four-epoch 15\,GHz VLBA data (see their Tables~1 and 2). To examine the jet structure on sub-parsec scales, we fitted a simple power-law function to these data. This yields a relationship of $W \propto r^{0.52\pm0.03}$ (reduced $\chi^{2}\approx 2.6$), indicative of a parabolic geometry for the upstream jet. Notably, the multifrequency core size closely follows the parabolic fit in the inner subparsec region. However, at $r \gtrsim 10^{4}\,R_{\rm s}$, the jet width appears to deviate from this trend, with the discrepancy becoming particularly pronounced on the kiloparsec scales when the VLA results from \citetalias{Nakahara_2018ApJ...854..148N} are included. To further investigate, we performed an additional power-law fit using jet width measurements from both VLBA and VLA, excluding the multifrequency core sizes. This fit yields $W \propto r^{0.97 \pm 0.01}$ (reduced $\chi^{2}\approx 3.2$), suggesting a conical geometry for the downstream jet.

These results reveal a parabolic-to-conical structural transition in the NGC\,4261 jet, occurring at distances of $\lesssim 10^{4}\,R_{\rm s}$, corresponding to sub-parsec scales of $\lesssim1.5$\,pc. As expected, both the deconvolution and the model fitting methods yield consistent results for the jet collimation (see Figures~\ref{fig: NGC4261_collimation_SLICE} and \ref{fig: NGC4261_collimation_MF}). However, since the former offers a significantly larger number of measurements (particularly for the counterjet) and achieves better statistical performance (i.e., lower reduced $\chi^2$; see Table~\ref{tab: NGC4261_broken_power-law_fit}), we adopt the deconvolution-based results as the basis for the primary conclusions of this study.

\subsection{Jet Velocity Field} \label{sec: Results_Jet_velocity_field}
Various methods can be used to study the kinematics and acceleration of AGN jets, as summarized by \citet{Park_kinematics_2019ApJ...887..147P} in their study of M\,87. Typically, detailed kinematic studies rely on long-term, extensive, and high-cadence monitoring of the jets. Unfortunately, due to the limited data available, this is currently not feasible for NGC\,4261. Therefore, we adopted alternative methods: (1) estimating the jet speed from the motion of local maxima in the radial intensity profile of the jet \citep[e.g.,][]{piner_2001AJ122.2954P}; (2) probing the jet kinematics using the jet-to-counterjet brightness ratio and spectral index \citep[e.g.,][]{park_2021ApJ...909...76P, Ricci_2022A&A...664A.166R}. In Appendix~\ref{Appendix: Jet_motions_derived_from_radial_intensity_profiles}, we provide further details on the use of the first method. In summary, our analysis suggests the possible intrinsic speeds of $(0.29 \pm 0.14)\,c$, $(0.50 \pm 0.24)\,c$, and $(0.56 \pm 0.14)\,c$ for the counterjet, at projected mean distances of 2.30, 5.70, and 5.72\,mas from the central engine, respectively (see Table~\ref{tab: NGC4261_jet_and_cjet_motions} in Appendix~\ref{Appendix: Jet_motions_derived_from_radial_intensity_profiles}). An intrinsic speed of $(0.37 \pm 0.04)\,c$ is also measured for the approaching jet at about 6.18\,mas. This corresponds to an apparent speed of $(0.80 \pm 0.15)\,\text{mas}\,\text{yr}^{-1}$, which is consistent with the value of $(0.83 \pm 0.11)\,\text{mas}\,\text{yr}^{-1}$ derived by \citet{piner_2001AJ122.2954P} using the same data set and method. In the following, we focus on using the second method to probe the jet kinematics and acceleration of NGC\,4261.

\subsubsection{Jet-to-Counterjet Brightness Ratio and Spectral Index Measurements} \label{sec: Results_measurements_of_BR_and_alpha}

\begin{figure*}[htbp!]
\begin{center}
    \includegraphics[width=0.4\linewidth]{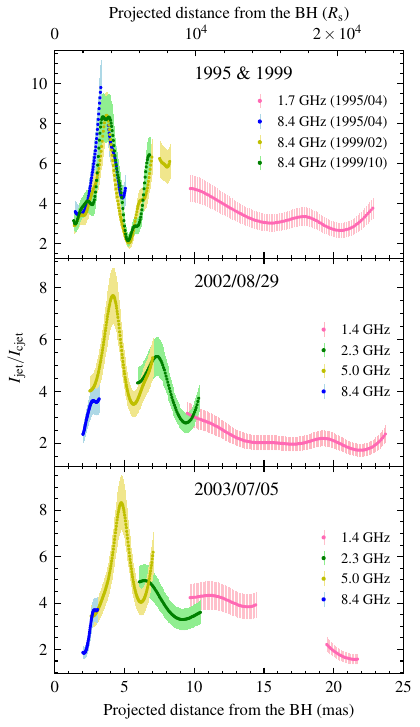}
    \includegraphics[width=0.4\linewidth]{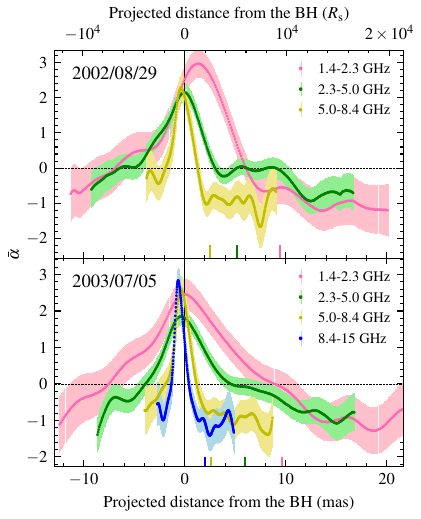}
\caption{Jet-to-counterjet brightness ratio (left) and spectral distribution of the twin jets (right) in NGC\,4261 as a function of projected distance from the central black hole. The colored solid sticks on the horizontal axes in the right panel indicate the regions (to the right of these sticks) where the brightness ratio are measured, in order to avoid the central blending, SSA, and FFA effects (see Figure~\ref{fig: NGC4261_jet_ridge_line}). 
\label{fig: NGC4261_BR_and_alpha}}
\end{center}
\end{figure*}

\begin{figure*}[htbp!]
\begin{center}
    \includegraphics[width=0.9\linewidth]{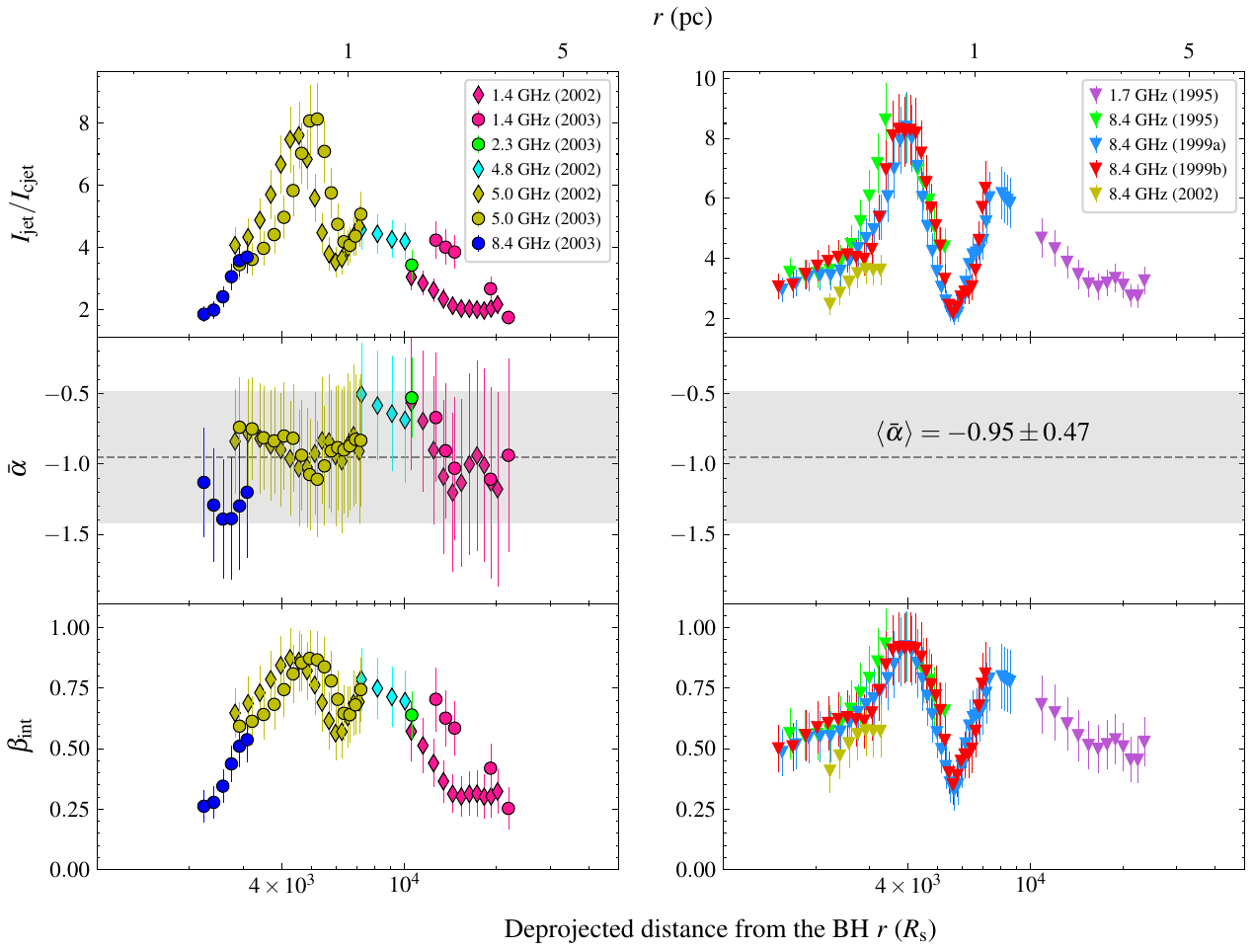}
\caption{Using the binned jet-to-counterjet brightness ratio (top) and spectral index (middle), the intrinsic speed of the NGC\,4261 jet (bottom) is derived.
In the middle plots, the gray dashed line and shaded area represent the average spectral index and its uncertainty, respectively. 
\textit{Left panels:} for the 2002 and 2003 data sets (except for the 8.4\,GHz data from 2002), the intrinsic speed is derived from each binned $R_{\rm B} - \bar{\alpha}$ pair. \textit{Right panels:} for data sets lacking spectral information, we adopt an average spectral index of $-0.95 \pm 0.47$ to derive the intrinsic speed.}
\label{fig: NGC4261_vint1}
\end{center}
\end{figure*}

The extended counterjet detected at frequencies $\leq$15\,GHz (see Figures~\ref{fig: NGC4261_jet_and_cjet_motions}, \ref{fig: NGC4261_1.4-8.4GHz_images}, and \ref{fig: NGC4261_15-43GHz_images}) allows us to measure the jet-to-counterjet brightness ratio ($R_{\rm B} = I_{\rm jet}/I_{\rm cjet}$, where $I_{\rm jet}$ and $I_{\rm cjet}$ are the observed intensity of the jet and counterjet, respectively). As illustrated in Figure~\ref{fig: NGC4261_jet_ridge_line}, we first shifted the putative location of the cental SMBH to the center of the image, based on the core shifts measured by \citet{haga2015ApJ80715H}. Next, we determined the maximum intensity along each slice perpendicular to the jet axis, excluding values smaller than 7$\sigma$ image rms noise (Figure~\ref{fig: NGC4261_jet_ridge_line}, left). This provided the radial intensity profile of the two-sided jets (Figure~\ref{fig: NGC4261_jet_ridge_line}, middle). The brightness ratio between the jet and counterjet was then calculated from the radial intensity profile (Figure~\ref{fig: NGC4261_jet_ridge_line}, right). The uncertainty in the brightness ratio was estimated via error propagation, assuming a 10\% uncertainty in intensity. Finally, given the complex effects of blending, synchrotron self-absorption (SSA), and FFA in the core region, we retained only measurements taken at distances from the observed core greater than the minor-axis beam size (see below).

We measured the spectral index ($\alpha$) by generating spectral index maps (SIMs) using quasi-simultaneously observed data from BS094b (1.4/2.3/5.0/8.4\,GHz) in 2002, BA064 (4.8/8.3\,GHz) in 2002, and BS094c/d (1.4/2.3/5.0/8.4/15\,GHz) in 2003 (see Table~\ref{tab: NGC4261_observation_summary}). SIMs were produced following the method detailed in \citet{Yan_2025NGC3998}. First, we excluded long-baseline data at higher frequencies and short-baseline data at lower frequencies to ensure comparable $(u,v)$ coverage for each frequency pair. We then restored new self-calibrated images for each pair, using the beam size of the lower-frequency image. After aligning the images considering the core shifts \citep{haga2015ApJ80715H}, we calculated the spectral index and its uncertainty at each pixel using Eqs.~1 and 2 from \citet{Yan_2025NGC3998}. Finally, the spectral distribution along the jet was derived by calculating the weighted average spectral index at each transverse slice (see Eq.~3 in \citealt{Yan_2025NGC3998}). The SIMs obtained at different frequency pairs are presented in Figure~\ref{fig: NGC4261_SIMs} of Appendix~\ref{Appendix: Spectral_index_maps}.

In Figure~\ref{fig: NGC4261_BR_and_alpha}, we show the jet-to-counterjet brightness ratio and spectral evolution of the twin jets in NGC\,4261 as a function of radial distance from the central black hole. As seen in the right panel, the jet base exhibits an optically thick, highly inverted spectrum ($\alpha > 2.5$), which is attributed to FFA by the foreground ionized gas \citep{Jones_1997ApJ...484..186J, Jones_2000ApJ...534..165J, Jones_2001ApJ...553..968J, haga2015ApJ80715H,Sawada-Satoh_2023PASJ...75..722S}. As mentioned earlier in this section, we excluded the inner FFA-affected region when measuring the jet-to-counterjet brightness ratio, which is indicated by the colored solid sticks on the horizontal axes in the plot. As seen, the extended jet on either side is characterized by an optically thin spectrum\footnote{However, the 2.3\,GHz data observed in 2002 do not show an optically thin spectrum until beyond 10\,mas in the approaching jet. Therefore, we exclude these data in the following.}. Therefore, the FFA and SSA effects, which are dominant near the jet base, become negligible, and the brightness difference between the jet and counterjet could be primarily affected by relativistic Doppler effects.

Interestingly, observations conducted at various epochs and frequencies reveal a consistent radial evolution of the jet-to-counterjet brightness ratio. As shown in the left panel of Figure~\ref{fig: NGC4261_BR_and_alpha}, the brightness ratio increases with distance in the inner region, peaking with values around $8-10$ at a projected distance of roughly $4-5$\,mas from the black hole. This is followed by a sharp decline in $R_{\rm B}$, then a subsequent rise. After reaching a second peak at about $7-8$\,mas from the black hole, $R_{\rm B}$ steadily decreases with increasing distance. Notably, this pattern appears to be independent of frequency, as long as the images at different frequencies capture a comparable region of the jet (see a similar finding obtained in NGC\,315 by \citealt{Ricci_2022A&A...664A.166R}). For instance, the two-epoch 5\,GHz data from 2002 and 2003, as well as the three-epoch 8.4\,GHz data from 1995 and 1999, independently reveal a similar evolutionary trend in $R_{\rm B}$. This consistency is likely because all these 5\,GHz and 8.4\,GHz images sample a nearly identical jet region, as shown in Figure~\ref{fig: NGC4261_jet_and_cjet_motions}. Similarly, at 1.4 and 1.7\,GHz, where a comparable jet morphology is detected (see Figure~\ref{fig: NGC4261_1.4-8.4GHz_images}), we observe a similar trend of a gradual decrease in $R_{\rm B}$ beyond $10$\,mas.

\subsubsection{Intrinsic Jet Speed} \label{sec: Results_Intrinsic_speed_of_the_jet}
Assuming that the two-sided jets in NGC\,4261 are intrinsically symmetric, the observed brightness difference between the jet and counterjet can be largely explained by relativistic Doppler boosting and de-boosting effects. Thus, using the jet-to-counterjet brightness ratio ($R_{\rm B}$), spectral index ($\alpha$), and jet viewing angle ($\phi_{\rm view} = 68^{\circ} \pm 4^{\circ}$; see Section~\ref{sec: Introduction}), we can calculate the intrinsic speed ($\beta_{\rm int}$) of the NGC\,4261 jet as follows:

\begin{equation} \label{eq: beta_int_1}
\beta_{\rm int} = \frac{1}{\cos\phi_{\rm view}}\left(\frac{R_{\rm B}^{1/(2-\alpha)}-1}{R_{\rm B}^{1/(2-\alpha)}+1}\right)
\end{equation}

For the 2002 and 2003 data sets, the jet-to-counterjet brightness ratio and spectral index were measured by a pixel-based analysis. As shown in Figure~\ref{fig: NGC4261_BR_and_alpha}, at each radial distance where $R_{\rm B}$ is measured, a corresponding weighted average spectral index is available, except for the 8.4\,GHz data from 2002. However, due to resolution limitations, these pixel-based measurements are not entirely independent. To address this, we binned $R_{\rm B}$ and $\bar\alpha$ by distance from the central engine, using a bin size of $\Theta_{\rm min}/5$. For each bin, we calculated the mean values for the radial distance, the brightness ratio, and the spectral index. In addition, we exclude binned $\bar\alpha$ values (and their corresponding $R_{\rm B}$) greater than $-0.5$. The intrinsic speed was then calculated for each binned $R_{\rm B} - \bar\alpha$ pair (see Eq.~\ref{eq: beta_int_1}), with uncertainties determined by error propagation. For data sets where spectral information is unavailable, we adopted an average spectral index of $-0.95 \pm 0.47$, derived from binned spectral indices as described above. The results are shown in Figure~\ref{fig: NGC4261_vint1}. As can be seen, the spectral index remains relatively stable across the explored distances, indicating that the pattern of the jet velocity field is primarily governed by variations in $R_{\rm B}$ (see Eq.~\ref{eq: beta_int_1}).

\begin{figure}[htbp!]
\begin{center}
    \includegraphics[width=1\linewidth]{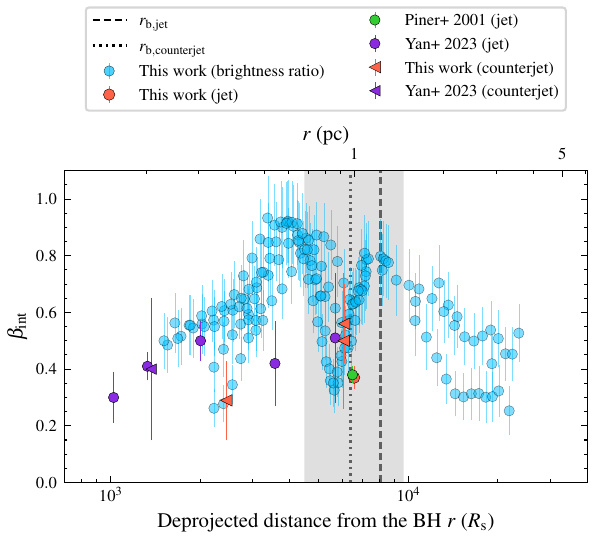}
\caption{Intrinsic speed of the NGC\,4261 jet as a function of deprojected distance from the central black hole. The blue points indicate the measurements derived using the jet-to-counterjet brightness ratio as proxy (see Figure~\ref{fig: NGC4261_vint1}). The orange symbols represent the intrinsic speeds derived from the jet’s radial intensity profile (see Appendix~\ref{Appendix: Jet_motions_derived_from_radial_intensity_profiles}). 
Green and violet symbols correspond to the measurements from \citet{piner_2001AJ122.2954P} and \citetalias{Yan_2023}, respectively. The vertical dashed and dotted lines indicate the break locations in the width profiles of the jet and counterjet, respectively, with the shaded region representing their 1$\sigma$ uncertainties (see Table~\ref{tab: NGC4261_broken_power-law_fit}). Note that the intrinsic speed of the counterjet is highlighted by the left-facing triangles (4 data points in total).}
\label{fig: NGC4261_vint2}
\end{center}
\end{figure}

Figure~\ref{fig: NGC4261_vint2} shows the intrinsic speed of the NGC\,4261 jet as a function of deprojected distance from the central black hole. In particular, the intrinsic speed derived from the jet’s radial intensity profile is also included (see Appendix~\ref{Appendix: Jet_motions_derived_from_radial_intensity_profiles}). Moreover, we incorporate kinematic results from \citetalias{Yan_2023}, based on four-epoch 15\,GHz VLBA data and the model fitting method. Their measured apparent speeds have been converted to intrinsic values using Eq.~\ref{eq: beta_int_2} in Appendix~\ref{Appendix: Jet_motions_derived_from_radial_intensity_profiles}. The core-shift effect has also been considered when determining the radial distances of the jet components relative to the black hole. \citet{piner_2001AJ122.2954P} have measured an apparent speed of $(0.83\pm0.11)$\,mas\,yr$^{-1}$ in the approaching jet, which is also presented in the figure.

Notably, measurements obtained through different methods, namely model fitting, radial intensity profiles, and the jet-to-counterjet brightness ratio, are broadly consistent. Together, they suggest an acceleration of the jet from $(0.30\pm0.09)\,c$ at $\sim10^3\,R_{\rm s}$ to the maximum relativistic speeds of $(0.92\pm0.15)\,c$ at $\sim4\times10^3\,R_{\rm s}$. The jet then undergoes a rapid deceleration, slowing to $(0.33\pm0.08)\,c$ at $\sim5.6\times10^3\,R_{\rm s}$, which is followed by a reacceleration with intrinsic speed up to $(0.81\pm0.13)\,c$ at $\sim8\times10^3\,R_{\rm s}$. Finally, the jet gradually decelerates to $(0.3-0.5)\,c$ at $(2-3)\times10^4\,R_{\rm s}$.

We note that the jet exhibits mildly complex kinematic evolution on scales of $(4-8)\times10^3\,R_{\rm s}$, characterized by a rapid deceleration followed by a reacceleration. This kinematic pattern appears to spatially coincide with the collimation transition region identified in Section~\ref{sec: Results_Deconvolved_jet_width}, $(6.4\pm1.9)\times10^3\,R_{\rm s} \lesssim r \lesssim (8.1\pm1.6)\times10^3\,R_{\rm s}$ (see Table~\ref{tab: NGC4261_broken_power-law_fit}), where both the approaching and receding jets transition from parabolic to conical shape. This coincidence may be indicative of relatively complicated dynamics or strong instabilities within the jet during the transition in its collimation profile.

Despite these local variations, the jet shows an overall accelerating trend over $(1-8)\times10^3\,R_{\rm s}$, while beyond $\sim 8\times10^3\,R_{\rm s}$, it undergoes a gradual deceleration from a relativistic speed of $(0.81\pm0.13)\,c$ to the sub-relativistic speeds of $(0.3-0.5)\,c$ at $(2-3)\times10^4\,R_{\rm s}$. Together with the collimation profile presented in Section~\ref{sec: Results_jet_collimation}, our findings suggest that acceleration and collimation of the NGC\,4261 jet could occur simultaneously on physical scales of $r \lesssim 10^{4}\,R_{\rm s}$, corresponding to sub-parsec scales of $r \lesssim 1.5$\,pc. A further discussion of this connection is presented in Section~\ref{sec: Discussion_Jet_acceleration}.

On the other hand, Figure~\ref{fig: NGC4261_vint2} may also hint at a stratification in the jet velocity field within the $(2-4)\times10^3\,R_{\rm s}$ range\footnote{However, this may also be attributed to the low S/N of the extended jet in the 8.4\,GHz data from 2002 and 2003 (see Figure~\ref{fig: NGC4261_jet_and_cjet_motions}, right panels).}. Moreover, a noticeable velocity deviation at a distance of around $3.5\times10^3\,R_{\rm s}$, as reported by \citetalias{Yan_2023}, cannot be ignored. This is likely because they were unable to detect rapid motions, due to the limited epochs and wide monitoring intervals of their four-epoch VLBA observations (see their Table~1).

\section{Discussion} \label{sec: Discussion}
\subsection{Jet Collimation} \label{sec: Discussion_Jet_collimation}
In this study, we performed an in-depth analysis of the width profiles for both the jet and counterjet in NGC\,4261 using multiple data analysis methods (see Figures~\ref{fig: NGC4261_collimation_SLICE} and \ref{fig: NGC4261_collimation_MF}). Our results reveal a clear structural transition from a parabolic to a conical shape on both sides, which aligns well with the previous findings by \citetalias{Nakahara_2018ApJ...854..148N} and \citetalias{Yan_2023}. Importantly, all studies suggest that this transition occurs on the sub-parsec scales of $\lesssim 1.5$\,pc (or $\lesssim 10^{4}\,R_{\rm s}$), which is substantially smaller than the Bondi radius of the central SMBH ($r_{\rm Bondi} \approx 99.2$\,pc). Moreover, the strong agreement between the width profiles of the jet and counterjet (see Figures~\ref{fig: NGC4261_collimation_SLICE}) also implies that jet properties are more likely to be regulated by the global state of the circumnuclear environment rather than local conditions (\citetalias{Nakahara_2018ApJ...854..148N}). In the following, we briefly discuss the physical mechanisms that may be responsible for jet collimation in NGC\,4261, based on studies in other AGNs.

\subsubsection{Jet Confinement on Sub-Parsec Scales ($\lesssim 1.5$ pc)} \label{sec: Discussion_Jet_confinement_on_parsec_scales}
The transition of the jet geometry from parabolic to conical shape in NGC\,4261 could be driven by a change in the external pressure profile \citep[e.g.,][]{Lyubarsky_2009ApJ...698.1570L}. Notably, this change should take place on sub-parsec scales. Following the discussion in \citetalias{Yan_2023}, we propose that the advection dominated accretion flow (ADAF), which extends to $\sim$ 0.5\,pc \citep[e.g.,][]{Nemmen_2014MNRAS.438.2804N, haga2015ApJ80715H}, could play a role in shaping the innermost parabolic jet structure in NGC\,4261. Alternatively, the disk wind could be another suitable candidate (e.g., \citealt{Yuan_2015ApJ...804..101Y, Globus_2016MNRAS.461.2605G, Nakamura_2018ApJ...868..146N, Blandford_2022MNRAS.514.5141B}; see \citealt{boccardi_NGC315_2021A&A...647A..67B} for more details), provided it does not extend too far from the central black hole.

Furthermore, the structural transition of AGN jets could also be linked to the initial magnetization at the jet base \citep[e.g.,][]{Kovalev_2020MNRAS.495.3576K}.
In this context, the initial jet magnetization parameter is likely to correlate with the maximum Lorentz factor \citep[e.g.,][]{Chen_2021ApJ...906..105C}. The relatively low $\Gamma_{\rm max} \approx 2.6$ observed in NGC\,4261 (see Section~\ref{sec: Results_Intrinsic_speed_of_the_jet}) tentatively suggests that its jet base may not be highly magnetized---similar to NGC\,315 \citep[$\Gamma_{\rm max} \approx 3$;][]{park_2021ApJ...909...76P}, but in contrast to M\,87 \citep[$\Gamma_{\rm max} \approx 10$;][]{Park_kinematics_2019ApJ...887..147P}.
This may help explain both the sub-parsec-scale collimation transition and the ``slow acceleration" of its jet (see Section~\ref{sec: Discussion_Slow_jet_acceleration}).

\subsubsection{Conical Expansion from Sub-Parsec to Kiloparsec Scales} \label{sec: Discussion_Conically_expanding}

Another key question concerns how the parabolic jet transitions to a conical expansion on sub-parsec scales. To understand this, jet kinematics beyond the collimation break may provide valuable insights. As shown in Figure~\ref{fig: NGC4261_vint2}, we observe progressive deceleration downstream of the transition zone in NGC\,4261. This behavior could be driven by in situ energy dissipation, where kinetic energy is converted into internal energy within the jet. As a result, the increased internal energy may cause the jet to be overpressured relative to the ambient medium, allowing it to expand more freely into the surrounding medium. Similar scenarios have been proposed for NGC\,6251 and NGC\,315 \citep{Tseng_2016ApJ...833..288T, park_2021ApJ...909...76P}.

However, the exact mechanism responsible for this energy conversion remains unclear. One possible explanation could be the formation of a recollimation shock in the transition zone, possibly triggered by a pressure mismatch between the jet and the ambient medium caused by the change in the external pressure profile (see Section~\ref{sec: Discussion_Jet_confinement_on_parsec_scales}) \citep{Gomez_1995ApJ...449L..19G, Mizuno_2015ApJ...809...38M, Fuentes_2018ApJ...860..121F}. As suggested by simulations, such shocks can induce strong instabilities within the jet, leading to its deceleration and heating, which thus facilitates the energy conversion process \citep[e.g.,][]{Mizuno_2015ApJ...809...38M, costa2025recollimation}. In this work, we did not find compelling evidence of a recollimation shock in NGC\,4261; however, Appendix~\ref{Appendix: Recollimation_Shock} presents a possible hint of its presence.

\begin{figure}[htbp!]
\centering
    \includegraphics[width=1\linewidth]{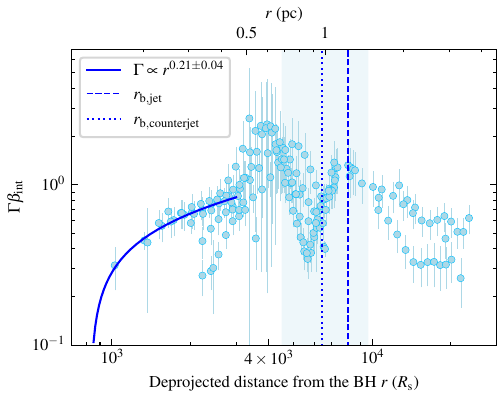}
\caption{Four-velocity of the NGC\,4261 jet as a function of deprojected distance from the central black hole, derived from the intrinsic speed as detailed in Section~\ref{sec: Results_Jet_velocity_field}. This plot includes the twin jets, with 4 data points corresponding to the counterjet (see Figure~\ref{fig: NGC4261_vint2}). The vertical dashed and dotted lines indicate the break locations in the width profiles of the jet and counterjet, respectively, with the shaded region representing their 1$\sigma$ uncertainties (see Table~\ref{tab: NGC4261_broken_power-law_fit}).}
\label{fig: NGC4261_gammaxbeta}
\end{figure}

\subsection{Jet Acceleration} \label{sec: Discussion_Jet_acceleration}
In Figure~\ref{fig: NGC4261_gammaxbeta}, we show the relationship between the four-velocity ($\Gamma\beta_{\rm int}$) of the NGC\,4261 jet and the deprojected distance from the central black hole. As mentioned in Section~\ref{sec: Results_Intrinsic_speed_of_the_jet}, although local kinematic variations are present, the overall kinematic trend indicates that the jet accelerates up to the jet collimation break at $(1.23\pm0.24)$\,pc or $(8.1\pm1.6)\times10^3\,R_{\rm s}$, beyond which it gradually decelerates. 
To examine the causal connection between the acceleration zone and the collimation zone, it is important to test whether the Lorentz factor and the jet opening angle ($\phi_{\rm open}$) satisfy the condition $\Gamma\phi_{\rm open}/2 \lesssim \sqrt{\sigma_{\rm m}}$, where $\sigma_{\rm m}$ is the magnetization parameter of the jet (i.e., Poynting-to-matter energy flux ratio) \citep{Komissarov_2009MNRAS.394.1182K}. For NGC\,4261, this condition appears to be fulfilled, with $\Gamma_{\rm max} \approx 2.6$, $\phi_{\rm open} \lesssim 5^{\circ}$, and an assumed $\sigma_{\rm m} \gtrsim 1$ within the ACZ. This suggests that efficient jet acceleration is taking place through magnetic energy conversion and that the acceleration and collimation processes in NGC\,4261 could be causally connected. Based on similar findings in M\,87, 1H 0323+342, NGC\,315, and 1928+738 (see Section~\ref{sec: Introduction}), we now identify a sub-parsec-scale ACZ ($\lesssim 1.5$\,pc) in the nearby LLAGN NGC\,4261.

\subsubsection{Slow Jet Acceleration} \label{sec: Discussion_Slow_jet_acceleration}
Assuming a simple power-law relationship between the Lorentz factor and the radial distance, we fitted a power-law function to $\Gamma$ within the range of $(1-3)\times10^3\,R_{\rm s}$, excluding data points that show significant deviations, as they may indicate potential stratification in the jet velocity field. Although a single power-law fit is somewhat simplistic, we obtained $\Gamma\propto r^{0.21\pm0.04}$ (see Figure~\ref{fig: NGC4261_gammaxbeta}). For comparison, the best-fit power-law functions in the jet acceleration zones of the two LLAGNs, M\,87 and NGC\,315, are $\Gamma\propto r^{0.16\pm0.01}$ and $\Gamma\propto r^{0.30\pm0.04}$, respectively \citep{Park_kinematics_2019ApJ...887..147P, park_2021ApJ...909...76P}. Similar to them, our results suggest a relatively shallow jet acceleration profile in NGC\,4261 when compared to predictions from highly magnetized jet models, which expect an efficient ``linear acceleration'' with $\Gamma \propto W \propto r^{k_{\rm u}} = r^{(0.48-0.51)}$ (see Table~\ref{tab: NGC4261_broken_power-law_fit}) in the inner parabolic jet region \citep{Tchekhovskoy_2008MNRAS.388..551T, Komissarov_2009MNRAS.394.1182K, Lyubarsky_2009ApJ...698.1570L}. In contrast, the FSRQ 1928+738 shows a steeper ``linear acceleration" profile, following $\Gamma \propto W \propto r^{0.46}$ \citep[][see a detailed discussion therein]{Yi_2024A&A...688A..94Y}.

\subsubsection{Jet Kinematics at the ACZ Boundary and Far Beyond} \label{sec: Discussion_Jet_Kinematics_at_the_ACZ_Boundary } 
As shown in Figures~\ref{fig: NGC4261_vint2} and \ref{fig: NGC4261_gammaxbeta}, the jet kinematics exhibits a degree of complexity prior to the end of the ACZ, featuring a rapid deceleration followed by reacceleration. This behavior may reflect intricate internal dynamics or instabilities associated with the transition in jet collimation (see Section~\ref{sec: Results_Intrinsic_speed_of_the_jet}). However, the physical mechanism underlying this behavior remains uncertain. One possible explanation involves the formation of a recollimation shock near the boundary of the ACZ (see Appendix~\ref{Appendix: Recollimation_Shock}), which may induce strong instabilities that affect the propagation of the jet \citep[e.g.,][]{Mizuno_2015ApJ...809...38M, costa2025recollimation}.

On the other hand, \citet{Laing_2014MNRAS.437.3405L} reported a flow speed of $\beta_{\rm int} = 0.92$ for the kiloparsec-scale jet in NGC\,4261. Given that the jet gradually decelerates to sub-relativistic speeds beyond approximately 1.5\,pc (see Figures~\ref{fig: NGC4261_vint2} and \ref{fig: NGC4261_gammaxbeta}), we propose that another acceleration mechanism should be at work on kiloparsec scales. Interestingly, \citet{park_2021ApJ...909...76P} also indicated the presence of an additional acceleration zone in the kiloparsec-scale, conically expanding jet in NGC\,315.

\begin{deluxetable*}{ccccc}[htbp!]
\renewcommand\arraystretch{1.3}
\tabletypesize{\footnotesize}
\tablecaption{Comparison of NGC\,4261 with M\,87 and NGC\,1052 }\label{tab: NGC4261_M87_NGC1052}
\tablehead{
& \colhead{NGC\,4261} 
& \colhead{M\,87$^a$} 
& \colhead{NGC\,1052$^b$} 
}
\startdata
Source distance ($D$)  & $\sim31.6$\,Mpc & $\sim16.8$\,Mpc & $\sim19.2$\,Mpc$^{(1)}$ \\
Black hole mass ($M_{\rm BH}$)  & $\sim1.62 \times 10^9\,M_{\odot}$ & $\sim6.5 \times 10^9\,M_{\odot}$ & $\sim1.6 \times 10^8\,M_{\odot}^{(2)}$ \\
Bondi radius ($r_{\rm Bondi}$)  & $\sim99.2$\,pc ($\sim6.5\times10^5\,R_{\rm s}$) & $\sim220$\,pc ($\sim3.5\times10^5\,R_{\rm s}$) & \ldots \\
Jet viewing angle ($\theta_{\rm view}$)   & $63^{\circ} - 71^{\circ}$ & $14^{\circ} - 20^{\circ}$  & $64^{\circ} - 87^{\circ}$\\
Jet width profile  & \makecell{parabolic $\to$ conical \\ (twin jets)}  
& \makecell{semi-parabolic $\to$ conical \\ (approaching jet)} 
& \makecell{parabolic $\to$ cylindrical $\to$ conical \\ (twin jets)} \\
Jet collimation break (JCB) & $r_{\rm break} \sim 1$\,pc $\ll r_{\rm Bondi}$ & $r_{\rm break} \sim r_{\rm Bondi}$ & \makecell{$r_{\rm break,1} \sim 0.05$\,pc \\ $r_{\rm break,2} \sim 0.15$\,pc}\\
Maximum Lorentz factor ($\Gamma_{\rm max}$) & $\Gamma_{\rm max} \approx 2.6$ at around JCB & $\Gamma_{\rm max} \approx 10$ at JCB & \ldots\\
Jet ACZ extent & $\sim1$\,pc $\ll r_{\rm Bondi}$ & $\sim220$\,pc $\sim r_{\rm Bondi}$  & \ldots\\
$\gamma$-ray emission detected?  & yes & yes & \ldots\\
Recollimation shock at the end of the ACZ?  & maybe & yes & \ldots \\
FFA in the counterjet?    & yes & no & yes$^{(3)}$ \\
Core ring diameter (1.3\,mm)  & $\sim5.6\,\mu$as$^{(4)}$  & $42\,\mu$as & ... \\
\enddata
\tablecomments{
References: (1) \citet{Tully_2013AJ....146...86T}; 
(2) \citet{Woo_2002ApJ...579..530W}; 
(3) e.g., \citet{Kameno_2001PASJ...53..169K, Kameno_2023ApJ...944..156K}, \citet{Vermeulen_2003AA...401..113V}, and \citet{Kadler_2004AA...426..481K}; 
(4) \citet{Zhang_2025ApJ98541Z}.
\flushleft $^{a}$ See Table~2 in \citet{Hada_2024} for a comprehensive summary of M\,87 and references therein.
$^{b}$ See references cited in Section~\ref{sec: Comparison with M 87 and NGC 1052}, unless otherwise noted. }
\end{deluxetable*}

\subsection{Comparison with M 87 and NGC 1052}  \label{sec: Comparison with M 87 and NGC 1052}
A brief comparison of the ACZs between NGC\,4261 and M\,87 (see also Table~\ref{tab: NGC4261_M87_NGC1052}):

\begin{itemize}[leftmargin=0.3cm, itemsep=0pt, topsep=1pt]
    \item[--] Both jets show a parabolic-to-conical structural transition. In M\,87, this transition occurs close to its Bondi radius \citep{Asada_2012ApJ...745L..28A}, whereas in NGC\,4261 the transition takes place only at $r\lesssim 1.5$\,pc, well within its Bondi radius.
    
    \item[--] Assuming a single power-law dependence of the Lorentz factor on the radial distance, both jets show a trend of gradual acceleration within their respective acceleration zones. For M\,87 this is characterized by $\Gamma\propto r^{0.16\pm0.01}$ \citep{Mertens_2016A&A...595A..54M,Park_kinematics_2019ApJ...887..147P}, while for NGC\,4261 it is $\Gamma\propto r^{0.21\pm0.04}$ (at distances of $1000R_{\rm s} - 3000R_{\rm s}$). Their maximum Lorentz factors are $\Gamma_{\rm max} \approx 10$ and $\Gamma_{\rm max} \approx 2.6$, respectively. 

    \item[--] These results suggest a co-spatial, causally connected ACZ in both LLAGNs. However, while the ACZ of M\,87 almost extends to its Bondi radius, the ACZ of NGC\,4261 is confined to a region smaller than 1.5\,pc, which is much less than its Bondi radius of $\sim99.2$\,pc.
    
    \item[--] High-energy $\gamma$-ray emission has been detected in these two low-luminosity radio galaxies. These emissions may originate from different jet regions, including the core/subparsec-scale jet and the larger hundred-parsec/kiloparsec-scale jet \citep[e.g.,][]{Aharonian_2006Sci...314.1424A,Cheung_2007ApJ...663L..65C,Harris_2009ApJ...699..305H,Abdo_2009ApJ...707...55A,Hada_2014ApJ...788..165H,Menezes_2020MNRAS.492.4120D,Tomar_2021ApJ...919..137T}. 

    \item[--] The recollimation shock has been clearly observed at the end of the ACZ in M\,87 \citep[i.e., HST-1; e.g.,][]{Stawarz_2006MNRAS.370..981S, Asada_2012ApJ...745L..28A}, while our results for NGC\,4261 also hint at a possible signature of recollimation at the boundary of its ACZ (see Appendix~\ref{Appendix: Recollimation_Shock}).

    \item[--] Based on the above comparisons, we tentatively suggest that the jet ACZ of NGC\,4261 is likely a scaled-down counterpart of that in M\,87. 
    
\end{itemize}

On the other hand, another nearby LLAGN, NGC\,1052, also deserves attention (see Table~\ref{tab: NGC4261_M87_NGC1052}):
\begin{itemize}[leftmargin=0.3cm, itemsep=0pt, topsep=1pt]
    \item[--] Similar to NGC\,4261, the NGC\,1052 jet is also inclined at a large viewing angle of $\phi_{\rm view} = 64^\circ-87^\circ$ \citep[e.g.,][]{Kadler_2004AA...426..481K,Baczko_2019A&A...623A..27B}. 

    \item[--] Interestingly, studies have suggested a nearly symmetric structural transition on both sides of NGC\,1052: a parabolic-to-cylindrical transition at around $3\times10^{3}\,R_{\rm s}$ and a cylindrical-to-conical transition at a distance of about $1\times10^{4}\,R_{\rm s}$ \citep{Nakahara_NGC1052_2020AJ....159...14N, Baczko_2022A&A...658A.119B, Baczko_2024A&A...692A.205B}. Notably, the second transition occurs on the subparsec scale of $\sim0.15$\,pc, which is comparable to the torus size of 0.5\,pc \citep{Kameno_2001PASJ...53..169K,Kameno_2003PASA...20..134K,Kameno_2020ApJ...895...73K,Kameno_2023ApJ...944..156K,Sawada-Satoh_2008ApJ...680..191S}. As discussed by \citet{Nakahara_NGC1052_2020AJ....159...14N} and \citet{Baczko_2022A&A...658A.119B}, the jet confinement in NGC\,1052 may be influenced by this dense, optically-thick torus. 
 
    \item[--] Kinematic studies in NGC\,1052 indicate that jet acceleration may occur at distances of $\lesssim10^{4}\,R_{\rm s}$, which is likely to coincide with its collimation process \citep{Baczko_2019A&A...623A..27B,Baczko_2022A&A...658A.119B}.
    
    \item[--] If this is the case, NGC\,1052 may host a subparsec-scale ACZ, similar to that observed in NGC\,4261 by this work.

    \item[--] NGC\,4261 and NGC\,1052 form a unique pair in which both the collimation profiles of the jet and counterjet have been measured in detail. Following the collimation studies in NGC\,1052, we propose that jet confinement in NGC\,4261 may also be influenced by a circumnuclear torus, although the physical parameters of this torus remain uncertain. 
\end{itemize}

\subsection{Caveats and Future Prospects}
\subsubsection{Observational Limitations}
It should be noted that NGC\,4261 lies near the celestial equator (declination $\delta\approx +6^{\circ}$), while the VLBA stations are distributed primarily along the east--west direction. This configuration results in poor sampling of north--south spatial frequencies in the $(u,v)$ coverage. 
Consequently, our results may be influenced by these overall image fidelity limitations.

\subsubsection{Limitations in Kinematic Analysis}
Notably, the archival data used in this study are constrained by limited sensitivity (see Table~\ref{tab: NGC4261_observation_summary}), resulting in insufficient S/N for the extended twin jets in some images (e.g., Figure~\ref{fig: NGC4261_jet_and_cjet_motions}, right). Low S/N can introduce significant uncertainties in position measurements and may also affect the accuracy of brightness ratio estimates for the faint, extended jet emission. In addition, the shape of the convolving beam --~whether circular, elliptical, or over-resolved~-- can influence jet parameters derived from image-plane measurements, such as the radial intensity profile, jet-to-counterjet brightness ratio, and spectral index. These effects, combined with low S/N, are key factors that could limit the accuracy of our velocity estimates. Finally, it should also be emphasized that our measurements are confined to a limited physical region of only $\sim(10^{3} - 2\times10^{4})\,R_{\rm s}$.

\subsubsection{Strategies to Mitigate Limitations}
To mitigate these limitations, it is important to (1) employ interferometric arrays with baseline configurations that offer improved $(u,v)$ coverage for NGC\,4261, and (2) adopt advanced imaging and kinematic analysis techniques. In future work, our aim is to further investigate the jet properties of NGC\,4261 using observational data from various arrays and frequencies, including the East Asia VLBI Network, the High Sensitivity Array, and the joint European VLBI Network with e-MERLIN. In parallel, we will apply advanced imaging techniques such as the Bayesian imaging method \citep{Kim_2024A&A...690A.129K,Kim_2025A&A...696A.169K} and the Regularized Maximum Likelihood method \citep[e.g.,][]{EHT_imaging_2019ApJ...875L...4E}, along with kinematic analysis techniques including the wavelet-based image segmentation and evaluation method \citep[e.g.,][]{Mertens_2016A&A...595A..54M}.

\subsubsection{Conclusion and Future Prospects}
With these caveats in mind, we conclude that this study provides the first robust evidence for a co-spatial sub-parsec ACZ in NGC\,4261. Following M\,87, 1H 0323+342, NGC\,315, and 1928+738 (see references in Section~\ref{sec: Introduction}), NGC\,4261 becomes the fifth radio-loud AGN (and the third LLAGN) to show compelling evidence of simultaneous jet acceleration and collimation.

Studying the ACZ properties of AGN jets is not only essential to understand the jets themselves, but also to unveil the mechanisms by which they are launched and collimated in the immediate vicinity of the central SMBHs, and how they propagate to larger kiloparsec scales. For NGC\,4261, to fully understand the formation of its sub-parsec ACZ, high-resolution imaging of the innermost jet regions, close to the SMBH, is indispensable. In this regard, the superior angular resolution enabled by millimeter-VLBI (mm-VLBI), combined with the development of multi-frequency receiving systems, will be crucial. Future observations with the Global mm-VLBI Array \citep{Ros_2024evn..conf..159R}, the ngEHT \citep{Doeleman_2023Galax..11..107D}, and the next-generation VLA \citep[ngVLA;][]{Murphy_2018ASPC..517....3M, Kadler_2023arXiv231110056K} will play an unique role in advancing our understanding of jet formation, acceleration, and collimation in this source. For example, what are the width \citep[cf. M\,87;][]{Kim_2018A&A...616A.188K,Lu_2023,Cui_2025NatAs.tmp..151C} and velocity profiles of the innermost jet down to a few tens of $R_{\rm s}$? As recently discovered in other nearby sources (see \citealt{Park_2024ApJ...973L..45P} and references therein), does the transverse jet structure exhibit a limb-brightened morphology?

Given the current limitations in ACZ studies, it is imperative to expand such studies to a larger AGN sample \citep[e.g.,][]{Boccardi_2025}. This will enable a comprehensive examination of the diversity in jet ACZ properties and facilitate statistical studies on how the ACZ extent correlates with central engine parameters (e.g., black hole spin, accretion mode, and magnetization) and external environments (e.g., disk winds) \citep[e.g.,][]{Okino_2022ApJ...940...65O,Fariyanto_2025arXiv250721241P}. Such efforts could shed light on the role of black hole, accretion mode, and ambient pressure in shaping jet dynamics, provide insights into the mechanisms driving the formation of relativistic jets \citep[e.g.,][]{Blandford_2019ARA&A..57..467B}, and ultimately enhance our understanding of AGN feedback and the coevolution of SMBHs with their host galaxies \citep[e.g.,][]{Kormendy_2013ARA&A..51..511K}.

\section{Summary} \label{sec: Summary}
We have studied the acceleration and collimation of the two-sided jets in NGC\,4261 using archival VLBA data, supplemented by results from the literature, covering a frequency range from 1.4 to 88\,GHz. Our main findings are summarized as follows.

\begin{enumerate}[leftmargin=0.4cm, itemsep=0pt]
    \item We conducted a detailed analysis of the jet collimation profile using multiple methods, including deriving the deconvolved jet width, modeling the jet structure with circular Gaussian functions, and examining the multifrequency properties of the VLBI core. The consistency between our results and previous studies (\citetalias{Nakahara_2018ApJ...854..148N,Yan_2023}) provides robust evidence for a structural transition from a parabolic to a conical geometry in both the jet and counterjet. In particular, this transition occurs at $(1.23\pm0.24)$\,pc $\approx (8.1\pm1.6)\times10^3\,R_{\rm s}$ for the jet and $(0.97\pm0.29)$\,pc $\approx (6.4\pm1.9)\times10^3\,R_{\rm s}$ for the counterjet, both of which are substantially smaller than the Bondi radius of the central SMBH ($r_{\rm Bondi} \approx 99.2$\,pc $\approx 6.5\times10^{5}\,R_{\rm s}$). This implies that external pressure could play a key role in jet confinement. The striking symmetry of the width profiles between the jet and counterjet (see Table~\ref{tab: NGC4261_broken_power-law_fit} and Figure~\ref{fig: NGC4261_collimation_SLICE}) further suggests that the jet collimation is more likely to be regulated by the global state of the circumnuclear environment rather than local conditions.

    \item Assuming that the observed brightness asymmetry between the jet and counterjet is primarily due to Doppler boosting and de-boosting effects, we derived the jet velocity field at distances from $\sim10^{3}$ to $\sim2\times10^{4}\,R_{\rm s}$ based on the jet-to-counterjet brightness ratio and spectral index. Supposing a simple power-law dependence of the Lorentz factor on the radial distance, we find $\Gamma \propto r^{0.21\pm0.04}$ within $(1-3)\times10^3\,R_{\rm s}$, indicative of a ``slow acceleration'' on these scales. At larger distances of $(4-8)\times10^3\,R_{\rm s}$, the jet shows a mildly complex kinematic behavior characterized by a rapid deceleration followed by reacceleration. We note that this region appears to be spatially coincident with the collimation transition zone, $(6.4\pm1.9)\times10^3\,R_{\rm s} \lesssim r \lesssim (8.1\pm1.6)\times10^3\,R_{\rm s}$, potentially reflecting the complicated internal dynamics and/or strong instabilities associated with the structural transition. Further studies are critical to explore the origin of such jet dynamics or instabilities (e.g., the formation of a recollimation shock at the collimation break zone; see Appendix~\ref{Appendix: Recollimation_Shock}) and to examine whether similar kinematic patterns are present in other AGNs.

    \item Despite these local kinematic variations, the jet undergoes an overall acceleration to relativistic speeds over the range $\sim(1-8)\times10^3\,R_{\rm s}$, with a maximum Lorentz factor of $\Gamma_{\rm max} \approx 2.6$. Beyond this region, it gradually decelerates to sub-relativistic speeds. The close spatial coincidence of the acceleration and collimation zones supports the existence of a co-spatial ACZ in NGC\,4261, which is notably confined to the sub-parsec scales of $r \lesssim 1.5$\,pc, corresponding to $r \lesssim 10^{4}\,R_{\rm s}$.

    \item We propose that within the sub-parsec-scale ACZ, the parabolic jet in NGC\,4261 may be confined by external pressure from the ADAF and disk wind (provided that the wind does not extend too far from the central SMBH), and the circumnuclear torus may also be a possible contributor. Meanwhile, the jet is gradually accelerated to relativistic speeds through the conversion of magnetic energy into kinetic energy. Furthermore, based on the maximum Lorentz factor of the jet, we tentatively suggest that the jet base in NGC\,4261 may not be highly magnetized, resembling NGC\,315 but differing from M\,87. This may qualitatively explain both the sub-parsec-scale structural transition and the observed ``slow acceleration'' in NGC\,4261.
    
\end{enumerate}

\begin{acknowledgments}
We sincerely thank the anonymous referee for the valuable comments which helped us greatly improve the quality of the manuscript. This work was supported by the CAS `Light of West China' Program (grant No. 2021-XBQNXZ-005) and the National SKA Program of China (grant No. 2022SKA0120102). 
X.Y. was supported by the China Postdoctoral Science Foundation under Grant Number 2025M773200. X.Y. also acknowledges support from the Xinjiang Tianchi Talent Program and the 2025 Outstanding Postdoctoral Grant of the Xinjiang Uygur Autonomous Region.
L.C. acknowledges the support from the Tianshan Talent Training Program (grant No. 2023TSYCCX0099). 
K.H. acknowledges the support from MEXT/JSPS KAKENHI (grant No. 25H00660, 22H00157, 21H04488) and the Mitsubishi Foundation (grant No. 202310034).
S.F. thanks the Hungarian National Research, Development and Innovation Office (NKFIH) for support (grants OTKA K134213 and TKP2021-NKTA-64). 
R.-S.L. is supported by the National Science Fund for Distinguished Young Scholars of China (grant No. 12325302) and the Shanghai Pilot Program for Basic Research, CAS, Shanghai Branch (JCYJ-SHFY-2021-013).
Chen.L. is supported by NSFC (12173066), National Key R\&D program of China (2024YFA1611403) and Shanghai Pilot Programme for Basic Research, CAS Shanghai Branch (JCYJ-SHFY-2021-013). 
L.C.H. was supported by the National Science Foundation of China (12233001) and the National Key R\&D Program of China (2022YFF0503401). L.C.H. acknowledges support from the Xinjiang Tianchi Talent Program.
This work was also partly supported by the Urumqi Nanshan Astronomy and Deep Space Exploration Observation and Research Station of Xinjiang (XJYWZ2303). 
The Very Long Baseline Array is operated by the National Radio Astronomy Observatory, a facility of the National Science Foundation operated under cooperative agreement by Associated Universities, Inc.
\end{acknowledgments}

\facilities{VLBA (NRAO).}
\software{{\tt AIPS} \citep{Greisen2003},  
          {\tt Astropy} \citep{Astropy_2013A&A},
          {\tt DIFMAP} \citep{Shepherd_Difmap_1997ASPC..125...77S},
          {\tt IPython} \citep{Perez_iPython_4160251},
          {\tt Jupyter} \citep{Kluyver_Jupyter},
          {\tt Matplotlib} \citep{Hunter_Matplotlib},
          {\tt Numpy} \citep{Walt_Numpy,Harris_Numpy},
          {\tt Pandas} \citep{Mckinney_Pandas},
          {\tt Python} \citep{Oliphant_python,Millman_python},
          {\tt Scipy} \citep{Virtanen_SciPy}.
          }

\bibliography{references} 
\bibliographystyle{aasjournal}

\begin{figure*}[ht!]
\begin{center}
    \includegraphics[width=0.31\linewidth]{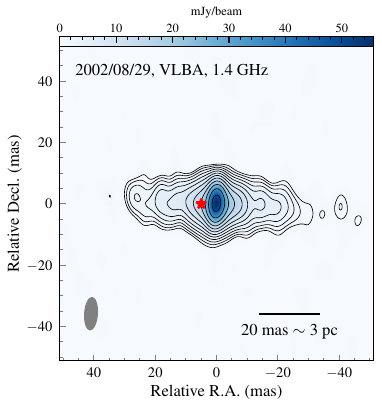}
    \includegraphics[width=0.31\linewidth]{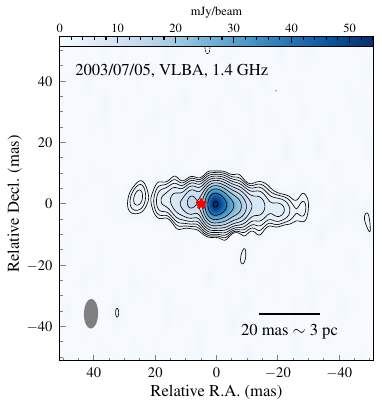}
    \includegraphics[width=0.31\linewidth]{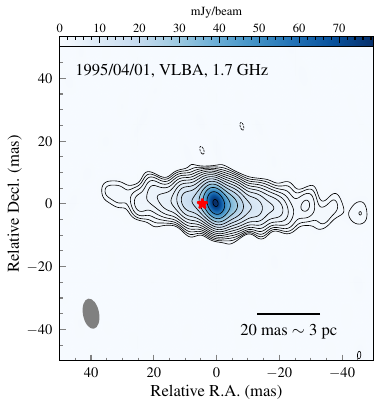}
    \includegraphics[width=0.32\linewidth]{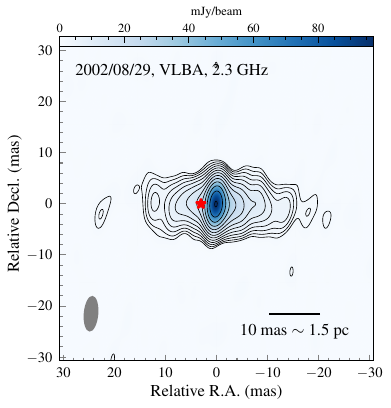}
    \hspace{-0.2cm}
    \includegraphics[width=0.32\linewidth]{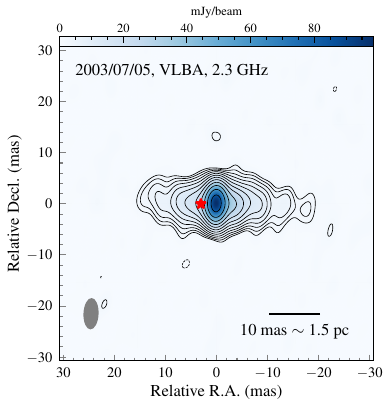}
    \hspace{-0.2cm}
    \includegraphics[width=0.32\linewidth]{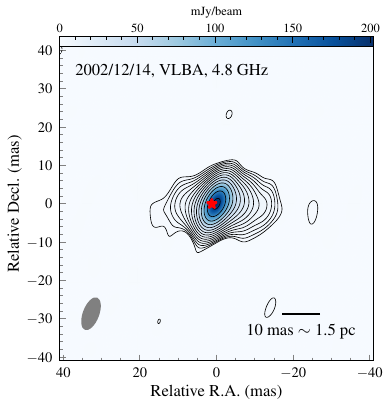}
    \hspace{-2cm}
    \includegraphics[width=0.31\linewidth]{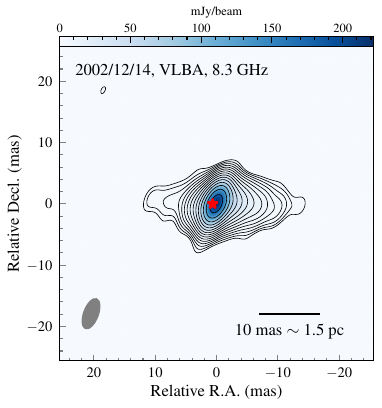}
    \includegraphics[width=0.31\linewidth]{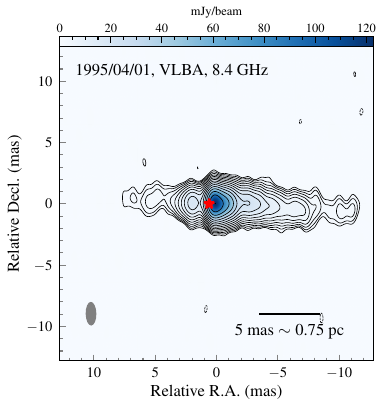}
\caption{Uniformly weighted CLEAN images of the NGC\,4261 jet observed with VLBA at 1.4, 1.7, 2.3, 4.8, 8.3, and 8.4\,GHz. Contours start from the 3$\sigma$ image rms noise (see Table~\ref{tab: NGC4261_observation_summary}) and increase by a factor of $\sqrt{2}$. The synthesized beam is shown in the bottom left corner of each image. The red star denotes the location of the black hole, determined based on the core shift measured by \citet{haga2015ApJ80715H} (see Table~\ref{tab: NGC4261_modelfitted_core_size}). Note that the 4.8 and 8.3\,GHz images observed in 2002 are affected by both limited sensitivity and resolution due to the short integration time and the limited number of antennas capable of detecting fringes (see Table~\ref{tab: NGC4261_observation_summary}).}
\label{fig: NGC4261_1.4-8.4GHz_images}
\end{center}
\end{figure*}

\begin{figure*}[ht!]
\begin{center}
    \includegraphics[width=0.3\linewidth]{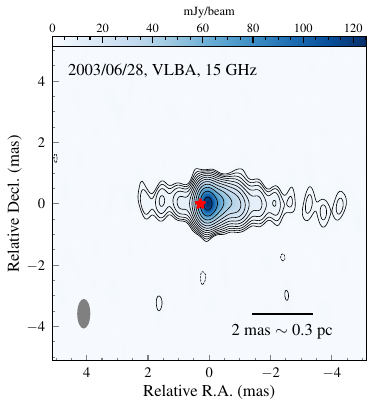}
    \includegraphics[width=0.305\linewidth]{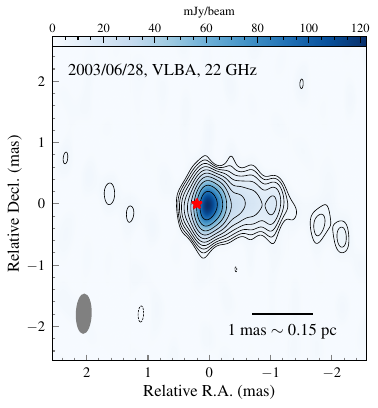}
    \includegraphics[width=0.325\linewidth]{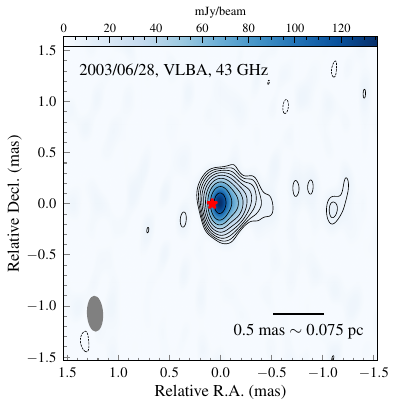}
\caption{Similar to Figure~\ref{fig: NGC4261_1.4-8.4GHz_images}, but observed at 15, 22, and 43\,GHz. 
\label{fig: NGC4261_15-43GHz_images}}
\end{center}
\end{figure*}

\begin{deluxetable}{cccccccccccc}[hb!]
\tablecaption{Model-fitted Multifrequency VLBI Core Properties of the Jet and Counterjet in NGC\,4261 \label{tab: NGC4261_modelfitted_core_size}}
\tablehead{\colhead{$\nu$} & \colhead{Epoch} & \multicolumn{4}{c}{Approaching jet core} & \multicolumn{4}{c}{Counterjet core}  & \colhead{Ref.} \\
\cmidrule(r){3-6}  \cmidrule(r){7-10} 
& & $S_{\nu}$ & $d$ & $\bar{d}$ & $r_{\rm to~BH}$ &  $S_{\nu}$ & $d$ & $\bar{d}$ & $r_{\rm to~BH}$ &   \\
 (GHz) & & (mJy) & (mas) & (mas)& (mas) & (mJy) & (mas) & (mas)& (mas) & \\
  (1) & (2) & (3) & (4) & (5) & (6) & (7) & (8) & (9) & (10) & (11)
}
\startdata
\multirow{2}{*}{1.4} & 2002/08/29 & $58\pm6$ & $2.47\pm0.62$ & \multirow{2}{*}{$2.67\pm0.67$} & \multirow{2}{*}{$5.002\pm2.06$} & $22\pm2$ & $6.48\pm1.62$ & \multirow{2}{*}{$5.27\pm1.32$} & \multirow{2}{*}{$4.82\pm2.05$}& \multirow{2}{*}{This work}  \\
                     & 2003/07/05 & $59\pm6$ &  $2.86\pm0.72$ & & & $20\pm2$ & $4.06\pm1.02$ \\
\hline
1.7 & 1995/04/01 & $78\pm8$ &  $1.48\pm0.37$ & $1.48\pm0.37$ & $4.524\pm1.89^{a}$ & $25\pm3$ & $3.87\pm0.97$ & $3.87\pm0.97$ & $4.53\pm1.96^{a}$ & This work \\
\hline
\multirow{2}{*}{2.3} & 2002/08/29 & $105\pm11$ &  $1.37\pm0.34$ & \multirow{2}{*}{$1.51\pm0.38$} & \multirow{2}{*}{$3.062\pm0.79$} & $22\pm2$ & $3.51\pm0.88$ & \multirow{2}{*}{$3.65\pm0.91$} & \multirow{2}{*}{$2.81\pm$0.79}  & \multirow{2}{*}{This work} \\
                     & 2003/07/05  &$109\pm11$ &  $1.65\pm0.41$ & & & $32\pm3$ & $3.78\pm0.95$   \\
\hline
\multirow{2}{*}{5.0} & 2002/08/29 & $126\pm13$ &  $0.86\pm0.22$ & \multirow{2}{*}{$0.84\pm0.22$} & \multirow{2}{*}{$1.262\pm0.17$} & $29\pm3$ & $1.51\pm0.38$ & \multirow{2}{*}{$1.57\pm0.39$}  & \multirow{2}{*}{$2.05\pm0.17$}  & \multirow{2}{*}{This work} \\
                     & 2003/07/05 & $136\pm14$ &  $0.82\pm0.21$ & & & $32\pm3$ & $1.62\pm0.41$   \\
\hline
\multirow{5}{*}{8.4} & 1995/04/01 & $113\pm11$ &  $0.39\pm0.10$ & \multirow{5}{*}{$0.45\pm0.12$}& \multirow{5}{*}{$0.602\pm0.08$}& $25\pm3$ & $0.77\pm0.19$ & \multirow{5}{*}{$1.28\pm0.32$} & \multirow{5}{*}{$1.68\pm0.08$} & \multirow{5}{*}{This work} \\
                     & 1999/02/26 & $80\pm8$   &   $0.28\pm0.07$  & & & $20\pm2$ & $0.63\pm0.16$ \\
                     & 1999/10/21 & $156\pm16$ &   $0.59\pm0.15$ & & & $32\pm3$ & $1.68\pm0.42$  \\
                     & 2002/08/29 & $167\pm17$ &   $0.51\pm0.13$ & & & $38\pm4$ & $1.84\pm0.46$  \\
                     & 2003/07/05 & $127\pm13$ &   $0.50\pm0.13$ & & & $35\pm4$ & $1.47\pm0.37$  \\
\hline
\multirow{5}{*}{15} & 2002/07/05 & $131\pm13$ &  $0.29\pm0.10$ & \multirow{5}{*}{$0.28\pm0.10$} & \multirow{5}{*}{$0.292\pm0.03$} & $30\pm3$ & $0.87\pm0.22$ & \multirow{5}{*}{$0.73\pm0.18$} & \multirow{5}{*}{$0.81\pm0.33$}  & \multirow{5}{*}{\makecell{\citetalias{Yan_2023}, \\ this work}} \\
                     & 2002/09/27 & $126\pm13$&  $0.27\pm0.10$ & & & $20\pm2$ & $0.89\pm0.22$ \\
                     & 2003/05/05 & $130\pm13$&  $0.26\pm0.09$ & & & $12\pm1$ & $0.62\pm0.16$  \\
                     & 2003/06/28 & $137\pm14$&  $0.28\pm0.09$ & & & $17\pm2$ & $0.63\pm0.16$  \\
                     & 2003/07/04 & $133\pm13$&  $0.26\pm0.09$ & & & $17\pm2$ & $0.66\pm0.17$  \\
\hline
22 & 2003/06/28 & $171\pm17$ &  $0.25\pm0.06$ & $0.25\pm0.06$ & $0.202\pm0.02$ & $12\pm1$ & $0.36\pm0.09$ & $0.36\pm0.09$ & $0.37\pm0.31$ & This work \\
\hline
43 & 2003/06/28 & $200\pm20$&  $0.17\pm0.04$ & \multirow{3}{*}{$0.17\pm0.04$}& \multirow{3}{*}{$0.082\pm0.02$} & ... & ... & \multirow{3}{*}{$0.28\pm0.07$} & \multirow{3}{*}{$0.25\pm0.03$}  & \multirow{3}{*}{\makecell{\citetalias{Yan_2023}, \\ this work}} \\
43                   & 2004/12/20 & $212\pm21$ &  $0.17\pm0.04$ & & & $22\pm2$ & $0.27\pm0.07$ & \\
44                   & 2022/02/14 & $166\pm17$ &  $0.17\pm0.04$ & & & $15\pm2$ & $0.28\pm0.07$ & \\
\hline    
88 & 2022/02/14 & $67\pm7$&  $0.09\pm0.02$ & $0.09\pm0.02$ & $0.036\pm0.02^a$ & ... & ... & ... & ... & \citetalias{Yan_2023} \\
\enddata
\tablecomments{
Column\,(1): frequency.  
Column\,(2): observation date.  
Columns\,(3)--(6): flux density, core size (FWHM), averaged core size derived from column\,(4), and projected radial distance relative to the central SMBH (adopted from \citealt{haga2015ApJ80715H}) for the approaching jet core. 
Columns\,(7)--(10): same as columns\,(3)--(6), but for the counterjet core.  
Column\,(11): references. 
\flushleft $^{a}$ Note that the core positions at 1.7 and 88\,GHz are estimated based on the core-shift relationship from \citet{haga2015ApJ80715H}.  
}
\vspace{-0.3cm}
\end{deluxetable}

\begin{deluxetable}{ccccccccccc}[hb!]
\tablecaption{Model-fitted Parameters of the Circular Gaussian Components Used to Derive the Jet Collimation Profile of NGC\,4261 \label{tab: NGC4261_modelfitted_jet_components}}
\tablehead{ \colhead{$\nu$}  & \colhead{Epoch} & \colhead{Comp.}  & \colhead{$S_{\nu}$} & \colhead{PA} & \colhead{$r_{\rm fit}$} & \colhead{$r_{\rm to\,core}$} & \colhead{$r_{\rm to\,BH}$}  & \colhead{$d$} & \colhead{$\Theta_{\rm min}/2$} & \colhead{Reduced $\chi^{2}$} \\
 (GHz) & & & (mJy) & (deg) & (mas) & (mas) & (mas) & (mas) & (mas) & \\
 (1) & (2) & (3) & (4) & (5) & (6) & (7) & (8) & (9) & (10) & (11)
}
\startdata
1.7 & 1995/04/01    & core & $78\pm8$ & 72.3 & $0.77$ & 0 & $4.524 \pm 1.89$ & $1.48\pm0.37$ & \multirow{8}{*}{2.57} & \multirow{8}{*}{0.96} \\
                    & & jet  & $8\pm1$  & $-92.1$ & 30.93 & 31.70 & $36.22\pm6.62$ &  $11.89\pm2.97$ &  \\
                    & &      & $12\pm1$ & $-93.7$ & 17.28 & 18.05 & $22.57\pm4.08$ & $5.20\pm1.30$ & \\
                    & &      & $16\pm2$ & $-94.1$ & 10.19 & 10.96 & $15.48\pm2.90$ & $3.09\pm0.77$ & \\
                    & &      & $36\pm4$ & $-93.1$ & 4.22  & 4.99 & $9.51\pm2.14$ & $2.08\pm0.52$ & \\
                    & & cjet & $25\pm3$ & 85.2    & 7.31 & 6.54 & $2.02\pm2.30$ & $3.87\pm0.97$ & \\
                    & &      & $10\pm1$ & 86.6    & 14.06  & 13.29 & $8.77\pm3.26$ & $4.94\pm1.23$ & \\
                    & &      & $8\pm1$  & 87.3    & 23.02 & 22.25 & $17.73\pm4.84$ & $7.79\pm1.95$ & \\
\hline
5.0 & 2002/08/29    & core & $126\pm13$  & 88.8 & 0.24& 0 & $1.262\pm0.17$ & $0.86\pm0.22$ & \multirow{7}{*}{0.63} & \multirow{7}{*}{1.15} \\
                    & & jet  & $16\pm2$ & $-$91.0 & 9.12 & 9.36 & $10.62\pm1.88$ & $4.03\pm1.01$ &\\
                    & &      & $26\pm3$ & $-$92.8 & 5.40 & 5.64 & $6.90\pm1.14$ & $2.45\pm0.61$ & \\
                    & &      & $43\pm4$ & $-$92.7 & 2.65  & 2.89 & $4.15\pm0.60$ & $1.72\pm0.43$ & \\
                    & &      & $50\pm5$ & $-$99.4 & 0.90 & 1.14 & $2.40\pm0.28$ & $1.10\pm0.27$ & \\
                    & & cjet & $29\pm3$ & 92.5  & 2.94 & 2.70 & $1.44\pm0.57$ & $1.51\pm0.38$ & \\
                    & &      & $12\pm1$ & 89.0 & 6.54 & 6.30 & $5.04\pm1.27$ & $3.12\pm0.78$ & \\
\hline
8.4 & 1995/04/01    & core & $113\pm11$  & 81.6 &0.25 & 0  & $0.602\pm0.08$ & $0.39\pm0.10$ & \multirow{12}{*}{0.43} & \multirow{12}{*}{1.23} \\
                    & & jet  & $10\pm1$ & $-93.9$ & 10.47 & 10.72 & $11.32\pm2.15$ & $2.44\pm0.61$ & \\
                    & &      & $8\pm1$  & $-94.3$ & 7.17 & 7.42 & $8.02\pm1.49$ & $1.07\pm0.27$ & \\
                    & &      & $11\pm1$ & $-93.9$ & 6.00  & 6.25 & $6.85\pm1.25$ & $1.03\pm0.26$ & \\
                    & &      & $13\pm1$ &$-92.7$  & 4.59 & 4.84 & $5.44\pm0.97$ & $0.75\pm0.19$ & \\
                    & &      & $26\pm3$ & $-90.7$ & 3.26 & 3.51 & $4.11\pm0.71$ & $0.91\pm0.23$ & \\
                    & &      & $30\pm3$ & $-88.4$ & 2.08 & 2.33 & $2.93\pm0.47$ & $0.88\pm0.22$ & \\
                    & &      & $59\pm6$ & $-88.6$ & 0.98 & 1.23 & $1.83\pm0.26$ & $0.54\pm0.14$ &  \\
                    & &      & $50\pm5$ & $-91.9$ & 0.34 & 0.59 & $1.19\pm0.14$ & $0.29\pm0.07$ &  \\
                    & & cjet & $25\pm3$ & 88.1 & 1.83 & 1.58 & $0.98\pm0.33$ & $0.77\pm0.19$ &  \\
                    & &      & $8\pm1$  & 87.7 &2.78 & 2.53 & $1.93\pm0.51$ & $0.79\pm0.20$ &  \\
                    & &      & $11\pm1$ & 86.3 & 5.55 & 5.30 & $4.70\pm1.06$ & $2.85\pm0.71$ &  \\
\hline
8.4 & 1999/02/26    & core & $80\pm8$ & 81.9 & 0.25 & 0 & $0.602\pm0.08$ & $0.28\pm0.07$ & \multirow{16}{*}{0.37} & \multirow{16}{*}{1.18} \\
                    &&  jet  & $4\pm1$ & $-94.7$ & 12.88 & 13.13 & $13.73\pm2.63$ & $2.47\pm0.62$ & \\
                    & &      & $5\pm1$ & $-93.3$ & 9.83  & 10.08 & $10.68\pm2.02$ & $1.73\pm0.43$ & \\
                    & &      & $7\pm1$ & $-93.9$ & 7.96 & 8.21 & $8.81\pm1.64$ & $1.06\pm0.27$ & \\
                    & &      & $7\pm1$ & $-92.4$ & 6.75 & 7.00 & $7.60\pm1.40$ & $0.99\pm0.25$ & \\
                    & &      & $10\pm1$ & $-91.7$ & 5.45 & 5.70 & $6.30\pm1.14$ & $0.86\pm0.21$ & \\
                    & &      & $8\pm1$ & $-90.3$ & 4.16 & 4.41 & $5.01\pm0.89$ & $0.75\pm0.19$ & \\
                    & &      & $19\pm2$ & $-91.2$ &  3.14 & 3.39 & $3.99\pm0.68$ & $0.80\pm0.20$ & \\
                    & &      & $27\pm3$ & $-91.9$ & 1.97  & 2.22& $2.82\pm0.45$ & $0.80\pm0.20$ & \\
                    & &      & $40\pm4$ & $-91.0$ & 0.99 & 1.24 & $1.84\pm0.26$ & $0.56\pm0.14$ & \\
                    & &      & $57\pm6$ & $-106.8$ & 0.24 & 0.49 & $1.09\pm0.13$ & $0.32\pm0.08$ & \\
                    & & cjet & $20\pm2$ & 88.4 & 1.71 & 1.46 & $0.86\pm0.30$ & $0.63\pm0.16$ & \\
                    & &      & $8\pm1$ & 87.0 & 2.68  & 2.43 & $1.83\pm0.49$ & $0.71\pm0.18$ & \\
                    & &      & $4\pm1$ & 88.7 & 3.88 & 3.63 & $3.03\pm0.73$ & $1.23\pm0.31$ & \\
                    & &      & $5\pm1$ & 84.4 & 5.75 & 5.50 & $4.90\pm1.10$ & $1.41\pm0.35$ & \\
                    & &      & $3\pm1$ & 85.2 & 7.95 & 7.70 & $7.10\pm1.54$ & $1.89\pm0.47$ & \\
\enddata
\tablecomments{
Column\,(1): frequency.
Column\,(2): observation date. 
Column\,(3): approaching jet core, jet component, or counterjet component.
Columns\,(4)--(6): flux density, PA, and radial distance relative to the coordinate origin of the component. 
Column\,(7): projected radial distance with respect to the core of the approaching jet.
Column\,(8): projected radial distance relative to central SMBH, taking into account the core shift measured by \citet{haga2015ApJ80715H} (see column (6) in Table~\ref{tab: NGC4261_modelfitted_core_size}).
Column\,(9): component size (FWHM).
Column\,(10): half of the minor-axis beam size. Components with FWHM smaller than $\Theta_{\rm min}/2$ are excluded when determining the jet collimation profile.
Column\,(11): reduced $\chi^{2}$ value of model fit.
}
\end{deluxetable}

\appendix

\section{CLEAN Images of NGC 4261} \label{Appendix: CLEAN_Images}
The main text presents the representative images of the NGC\,4261 jet (see Figure~\ref{fig: NGC4261_jet_and_cjet_motions}). Here, we provide the remaining images in Figures~\ref{fig: NGC4261_1.4-8.4GHz_images} and \ref{fig: NGC4261_15-43GHz_images}.

\section{Parameters of Gaussian Components Fitted to the Core and Twin Jets} \label{Appendix: Tables}

In Table~\ref{tab: NGC4261_modelfitted_core_size}, we present the model-fitted multifrequency properties of the VLBI core for the jet and counterjet in NGC\,4261, based on both our archival data in Table~\ref{tab: NGC4261_observation_summary} and previous data/results from the literature (\citealt{haga2015ApJ80715H}; \citetalias{Yan_2023}). 
In Table~\ref{tab: NGC4261_modelfitted_jet_components}, we summarize the data and model-fitted parameters of the circular Gaussian components fitted to the twin jets of NGC\,4261. These results are used to derive the jet collimation profile of NGC\,4261 (see Section~\ref{sec: Results_Model-fitted_jet_width} and Figure~\ref{fig: NGC4261_collimation_MF}).

\section{Proper Motions of the Twin Jets Derived from Radial Intensity Profiles} \label{Appendix: Jet_motions_derived_from_radial_intensity_profiles}
Using two-epoch 8.4\,GHz data from project BJ028 (see Table~\ref{tab: NGC4261_observation_summary}), \citet{piner_2001AJ122.2954P} measured an apparent speed in the NGC\,4261 jet from its radial intensity profile. Following their approach, we further investigated the motions of the twin jets using a slightly larger data set, as shown in the top panel of Figure~\ref{fig: NGC4261_jet_and_cjet_motions}. The details of determining the radial intensity profile of the jet are described in Section~\ref{sec: Results_measurements_of_BR_and_alpha} (see Figure~\ref{fig: NGC4261_jet_ridge_line}). Our final results are shown in the bottom panel of Figure~\ref{fig: NGC4261_jet_and_cjet_motions}. Note that all measurements are taken with S/Ns $\geq 10$.

As seen, all profiles exhibit two peaks and a dip in the inner $\pm$2\,mas region, which appear to be nearly stationary over the $8-10$ month period. Furthermore, we observed three features in the counterjet, including two local maxima and a signature of expansion in the intensity profile, as indicated in each plot of Figure~\ref{fig: NGC4261_jet_and_cjet_motions} (bottom panel). These features suggest possible motion of the counterjet, with radial displacements of about $(0.40\pm0.08)$, $(0.66\pm0.13)$, and $(0.56\pm0.11)$\,mas at distances of approximately 2.30, 5.70, and 5.72\,mas from the central engine (see Table~\ref{tab: NGC4261_jet_and_cjet_motions}). These measurements correspond to apparent speeds of about $(0.47\pm0.16)$, $(0.78\pm0.27)$, and $(0.86\pm0.15)$\,mas\,yr$^{-1}$ (assuming an uncertainty of $\Theta_{\rm min}/5$), respectively. Furthermore, two well-defined local maxima are clearly visible in the approaching jet, observed in 1999, suggesting an apparent speed of $(0.80\pm0.15)$\,mas\,yr$^{-1}$. This is consistent with the value reported by \citet{piner_2001AJ122.2954P}, $(0.83\pm0.11)$\,mas\,yr$^{-1}$, who obtained their result using the same data set and method.

\begin{deluxetable}{ccccccccc}[htbp!]
\tablecaption{Possible Motions of the Jet and Counterjet in NGC\,4261 Derived from the Radial Intensity Profile \label{tab: NGC4261_jet_and_cjet_motions}}
\tablehead{\colhead{} & \colhead{$\nu$} & \colhead{Epochs} & \colhead{$\Delta r$}  & \colhead{$\Delta{t}$} & \colhead{$\mu_{r}$} & \colhead{$\beta_{\rm app}$}  &  \colhead{$\beta_{\rm int}$} & \colhead{$\bar{r}$} \\
 & (GHz) & & (mas) & (yr) & (mas\,yr$^{-1}$) & & & (mas) \\
& (1) & (2) & (3) & (4) & (5)& (6) & (7) & (8)
}
\startdata
Jet & 8.4 & 1999/02 $-$ 1999/10 & $0.52\pm0.10$ & 0.65 & $0.80\pm0.15$ & $0.40\pm0.07$ & $0.37\pm0.04$ & 6.18\\
\hline
\multirow{3}{*}{Counterjet}  
& 5.0 & 2002/08 $-$ 2003/07 & $0.66\pm0.13$ & 0.85 & $0.78\pm0.27$ & $0.39\pm0.13$ & $0.50\pm0.24$ & 5.70\\
& 8.4 & 1999/02 $-$ 1999/10 & $0.56\pm0.11$ & 0.65 & $0.86\pm0.15$ & $0.43\pm0.07$ & $0.56\pm0.14$ & 5.72\\
& 8.4 & 2002/08 $-$ 2003/07 & $0.40\pm0.08$ & 0.85 & $0.47\pm0.16$ & $0.24\pm0.08$ & $0.29\pm0.14$ & 2.30\\
\enddata
\tablecomments{
Column\,(1): frequency.
Column\,(2): observation dates.
Column\,(3): projected radial displacement of the local maxima.
Column\,(4): time separation.
Column\,(5): proper motion derived from the radial intensity profile.
Columns\,(6) -- (7): apparent speed and intrinsic speed in units of $c$, respectively.
Column\,(8): projected mean distance with respect to the central engine.
}
\vspace{-1cm}
\end{deluxetable}

Assuming an intrinsically symmetric jet and adopting a jet viewing angle of $\phi_{\rm view} = 68^{\circ}\pm 4^{\circ}$ (see Section~\ref{sec: Introduction}), the intrinsic speed can be calculated as follows:

\begin{equation} \label{eq: beta_int_2}
\beta_{\rm int, jet/cjet} = \frac{\beta_{\rm app}}{\sin\phi_{\rm view} \pm \beta_{\rm app} \cos\phi_{\rm view}},
\end{equation}
where $\beta_{\rm int}$ and $\beta_{\rm app}$ are the intrinsic and apparent speeds of the jet/counterjet, in units of $c$. The results are summarized in Table~\ref{tab: NGC4261_jet_and_cjet_motions}. We note that these measurements are constrained by the limited observing epochs. However, they may still be useful when compared with those obtained through other methods (see Section~\ref{sec: Results_Intrinsic_speed_of_the_jet}).

\section{Spectral Index Maps} \label{Appendix: Spectral_index_maps}
\begin{figure}[htbp!]
\begin{center}
    \includegraphics[width=0.32\linewidth]{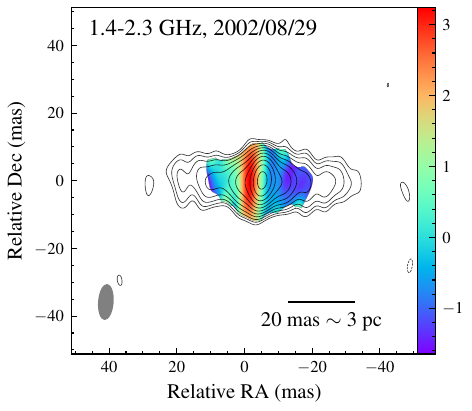}
    \includegraphics[width=0.32\linewidth]{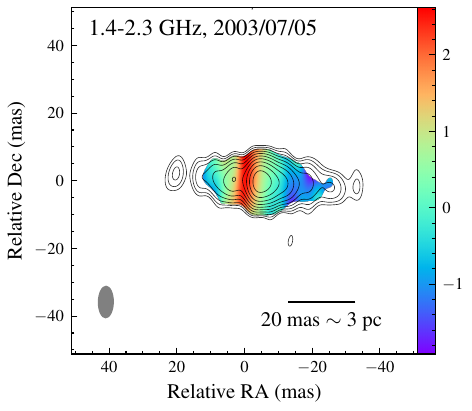}
    \includegraphics[width=0.33\linewidth]{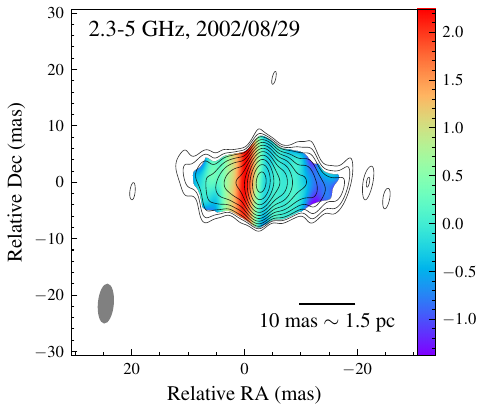}
    \includegraphics[width=0.33\linewidth]{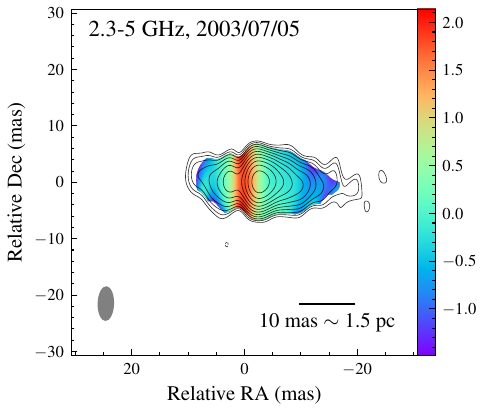}
    \includegraphics[width=0.32\linewidth]{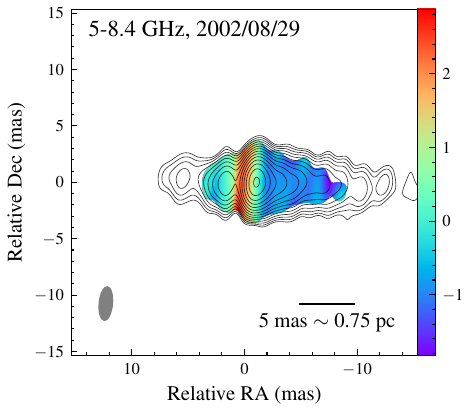}
    \includegraphics[width=0.32\linewidth]{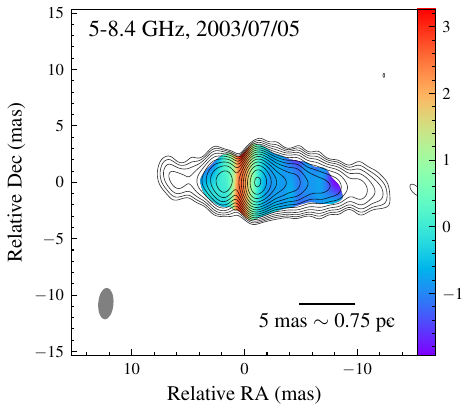}
    \includegraphics[width=0.35\linewidth]{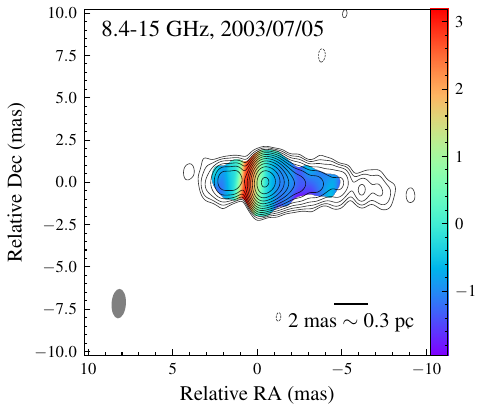}
    \includegraphics[width=0.35\linewidth]{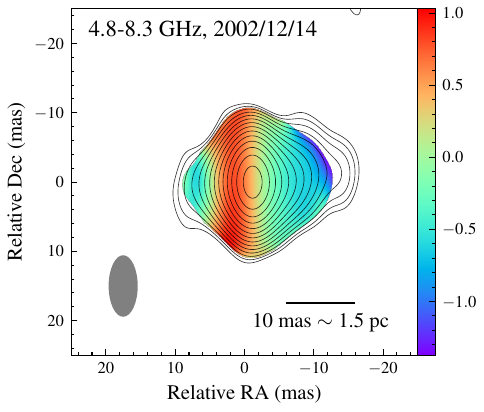}
\caption{SIMs at different frequency pairs of NGC\,4261. For each pair, the SIM is overlaid on the contours of the lower-frequency image. Note that the coordinate origin is set to the putative location of the central SMBH using the core shifts measured by \citet{haga2015ApJ80715H}.
\label{fig: NGC4261_SIMs}}
\end{center}
\end{figure}

As described in Section~\ref{sec: Results_measurements_of_BR_and_alpha} we generated SIMs to extract the spectral information of the twin jets, using quasi-simultaneously observed archival data. Here, we present the SIMs obtained from these data sets in Figure~\ref{fig: NGC4261_SIMs}. 
The highly inverted spectrum observed around the emission gap at the jet base, with $\alpha > 2.5$ in some cases, is attributed to FFA by the foreground ionized gas from the accretion disk \citep{Jones_1997ApJ...484..186J,Jones_2000ApJ...534..165J,Jones_2001ApJ...553..968J,haga2015ApJ80715H,Sawada-Satoh_2023PASJ...75..722S}.

\section{Possible Hint of a Recollimation Shock at the ACZ End of NGC\,4261} \label{Appendix: Recollimation_Shock}
The formation of a recollimation shock at the end of the ACZ has been proposed in several sources, including BL Lacertae \citep{Marscher_2008Natur.452..966M}, PKS 1510-089 \citep{Marscher_2010ApJ...710L.126M}, and OJ 287 \citep{Agudo_2011ApJ...726L..13A}. In particular, clear evidence for such shocks has been found in the jet collimation profiles of M\,87 \citep[e.g.,][]{Stawarz_2006MNRAS.370..981S, Asada_2012ApJ...745L..28A} and 1H 0323+342 \citep{Doi_2018ApJ...857L...6D, Hada_1H0323_2018ApJ...860..141H}. Specifically, an evident dip in jet width is seen near the location of the structural break, where the observed width is smaller than expected from a parabolic profile.

\begin{figure}[htbp!]
\begin{center}
    \includegraphics[width=0.5\linewidth]{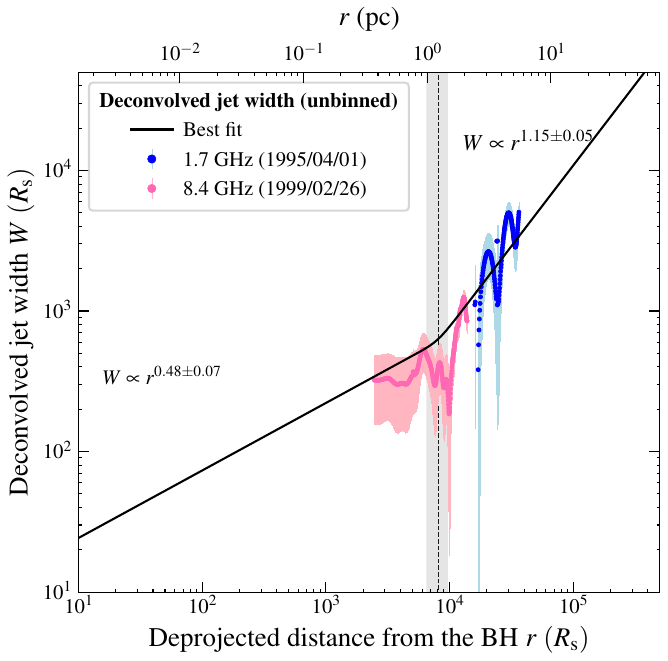}
\caption{Unbinned width profile of the approaching jet in NGC\,4261, derived from the 8.4\,GHz data (1999/02/26) and 1.7\,GHz data (1995/04/01). A dip is seen in the structural transition zone, potentially indicating a recollimation feature.}
\label{fig: NGC4261_recollimation_shock}
\end{center}
\end{figure}

For NGC\,4261, as shown in Figure~\ref{fig: NGC4261_recollimation_shock}, we identify a dip in the unbinned deconvolved width profile, enabled by the high-quality 8.4\,GHz data from 1999 February 26 (see Table~\ref{tab: NGC4261_observation_summary}). This feature may hint at the presence of a recollimation shock in the structural transition region, although confirmation will require higher-resolution observations.

\end{document}